\documentclass[longauth]{aa}  
\usepackage{graphicx}
\usepackage{natbib}
\bibpunct{(}{)}{;}{a}{}{,} 
\usepackage[varg]{txfonts}
\usepackage{hyperref}

\usepackage{color}
\usepackage{amsmath}
\usepackage{amssymb}
\usepackage{multicol}   
\usepackage{bm}         
\usepackage{pdflscape}  

\newcommand{\snr} {\mbox{S/N}}

\newcommand{\prot}{\mbox{P$_{\rm rot}$}}

\newcommand{\ie}{i.\,e.}

\newcommand{\eg}{e.\,g.}

\newcommand{\teff}{$T_{{\rm eff}}$}
\newcommand{\kms}{\mbox{km\,s$^{-1}$}}
\newcommand{\ms}{\mbox{m s$^{-1}$}}

\newcommand{\vsini} {\mbox{$v \sin i_\star$}}

\newcommand{\vmacro}{$v_{\rm mac}$}
\newcommand{\vmicro}{\mbox{$\xi_{\rm t}$}}
\newcommand{\gfeh}{\mbox{$[{\rm Fe}/{\rm H}]$}}

\newcommand{\etal}{\mbox{\rm et al.}}

\newcommand{\logg}{\mbox{log~{\it g}}}
\newcommand{\loggf}{\mbox{log~{\it gf}}}

\newcommand{\mplanet}{\mbox{M$_{\rm p}$}}

\newcommand{\msun}{\mbox{M$_\odot$}}
\newcommand{\rsun}{\mbox{R$_\odot$}}
\newcommand{\mstar}{\mbox{M$_\star$}}
\newcommand{\rstar}{\mbox{R$_\star$}}
\newcommand{\istar}{\mbox{$i_\star$}}
\newcommand{\mjup}{\mbox{M$_{\rm J}$}}
\newcommand{\rjup}{\mbox{R$_{\rm J}$}}

\newcommand{\logRHK}{\mbox{$\log {\rm R}^{\prime}_{\rm HK}$}}
\newcommand{\halpha}{\mbox{H$\alpha$}}

\begin{document}

\title{The GAPS programme with HARPS-N at TNG}
\subtitle{XI. Pr\,0211 in M\,44: the first multi-planet system in an open cluster \thanks{Based 
on data obtained with (\textit{i}) the HARPS-N spectrograph on the Italian Telescopio Nazionale Galileo (TNG), operated on the island of La Palma by the INAF 
- Fundacion Galileo Galilei (Spanish Observatory of Roque de los Muchachos of the IAC); (\textit{ii}) the Tillinghast Reflector Echelle Spectrograph (TRES) on the
 1.5-meter Tillinghast telescope, located at the Smithsonian Astrophysical Observatory's Fred L. Whipple Observatory on Mt. Hopkins in Arizona; (\textit{iii}) the STELLA robotic telescopes in Tenerife, an AIP facility jointly operated by AIP and IAC. 
Figure~\ref{fig:actindex_RVs}, \ref{fig:asyindex}, \ref{fig:actindex}, \ref{fig:phot_rv_act_P1}, \ref{fig:phot_rv_act_P2} and \ref{fig:rv_residuals_GLS_k4}, and Table~\ref{table:rho_m} are made available in the online version of the paper. 
Table~\ref{table:linelist}, \ref{table:RV_table}, \ref{table:asymmetry_table} and \ref{table:activity_table}  are made available on-line at the CDS (\url{http://cdsweb.u-strasbg.fr/}).}}

\author{
L. Malavolta  \inst{1,2},
V. Nascimbeni  \inst{2},
G. Piotto  \inst{1,2},
S. N. Quinn  \inst{3},
L. Borsato \inst{1,2},
V. Granata \inst{1,2},
A.S. Bonomo \inst{6},
F. Marzari \inst{1},
L.R. Bedin  \inst{1,2},
M. Rainer \inst{5},
S. Desidera  \inst{2},
A.F. Lanza  \inst{4},
E. Poretti  \inst{5},
A. Sozzetti \inst{6},
R. J. White  \inst{3},
D. W. Latham \inst{7},
A. Cunial  \inst{1,2},
M. Libralato  \inst{1,2},
D. Nardiello  \inst{1,2},
C. Boccato \inst{2},
R.U. Claudi \inst{2},
R. Cosentino \inst{7,8},
E. Covino \inst{9},
R. Gratton  \inst{2},
A. Maggio \inst{10},
G. Micela \inst{10},
E. Molinari \inst{9,11},
I. Pagano \inst{4},
R. Smareglia  \inst{12},
L. Affer \inst{10},
G. Andreuzzi  \inst{8,13},
A. Aparicio \inst{14,15},
S. Benatti  \inst{2},
A. Bignamini   \inst{12},
F. Borsa \inst{5},
M. Damasso \inst{6},
L. Di Fabrizio \inst{8},
A. Harutyunyan \inst{8},
M. Esposito \inst{9},
A.F.M. Fiorenzano \inst{8},
D. Gandolfi \inst{16,17},
P. Giacobbe \inst{6},
J.I. González Hernández  \inst{14,15},
J. Maldonado \inst{10},
S. Masiero \inst{2},
M. Molinaro \inst{12},
M. Pedani \inst{8},
G. Scandariato \inst{4}
}

\institute{ 
Dipartimento di Fisica e Astronomia Galileo Galilei -- Universit\`a di Padova, Vicolo dell'Osservatorio 2, I-35122, Padova, Italy \\
E-mail: \texttt{luca.malavolta@unipd.it}
\and INAF -- Osservatorio Astronomico di Padova,  Vicolo dell'Osservatorio 5, I-35122, Padova, Italy
\and Department of Physics \& Astronomy -- Georgia State University, Atlanta, GA 30303, USA
\and INAF -- Osservatorio Astrofisico di Catania, Via S.Sofia 78, I-95123, Catania, Italy
\and INAF -- Osservatorio Astronomico di Brera, Via E. Bianchi 46, I-23807 Merate (LC), Italy
\and INAF -- Osservatorio Astrofisico di Torino, Via Osservatorio 20, I-10025, Pino Torinese, Italy.
\and Harvard-Smithsonian Center for Astrophysics, Cambridge, MA 02138 USA
\and Fundaci\'on Galileo Galilei - INAF, Rambla Jos\'e Ana Fernandez P\'erez 7, E-38712 Bre\~na
 Baja, TF - Spain
\and INAF -- Osservatorio Astronomico di Capodimonte, Salita Moiariello 16, I-80131, Napoli, Italy
\and INAF -- Osservatorio Astronomico di Palermo, Piazza del Parlamento, 1, I-90134, Palermo, Italy
\and INAF - IASF Milano, via Bassini 15, I-20133 Milano, Italy
\and INAF -- Osservatorio Astronomico di Trieste, via Tiepolo 11, I-34143 Trieste, Italy
\and INAF - Osservatorio Astronomico di Roma - Sede di Monteporzio Catone
Via di Frascati, 33 00040 Monte Porzio Catone
\and Instituto de Astrofísica de Canarias, Calle vía Láctea s/n, 38205 La Laguna, Tenerife, Spain
\and Departamento de Astrofísica, Universidad de La Laguna, 38200 La Laguna, Tenerife, Spain
\and Dipartimento di Fisica, Universit\'a di Torino, via P. Giuria 1,
I-10125, Torino, Italy
\and Landessternwarte K\"onigstuhl, Zentrum f\"ur Astronomie der
Universit\"at Heidelberg, K\"onigstuhl 12, D-69117 Heidelberg, Germany
}

\date{Received / Accepted}
\abstract{
Open cluster (OC) stars share the same age and metallicity, and, in general, their age and mass can be estimated with higher precision than for field stars. For this reason, OCs are considered an important laboratory to study the relation between 
the physical properties of the planets and those of their host stars, and the evolution of planetary systems.
However, only a handful of planets have been discovered around OC main-sequence stars so far, all of them in single-planet systems. For this reason we started  an observational campaign within the GAPS collaboration to search for and characterize planets in OCs}
{We monitored the Praesepe member Pr\,0211 to improve our knowledge of the eccentricity of the hot Jupiter (HJ) that is already known to orbit this star and search for additional intermediate-mass planets. An eccentric orbit for the HJ would support a  planet-planet scattering process rather than a disk-driven migration after its formation.}
{From 2012 to 2015, we collected 70 radial velocity (RV) measurements with HARPS-N  and 36 with TRES of Pr\,0211. Simultaneous photometric observations were carried out with the robotic STELLA telescope to characterize the stellar activity. We discovered a long-term trend in the RV residuals that we show as being due to the presence of a second, massive, outer planet. Orbital parameters for the two planets are derived by simultaneously fitting RVs and photometric light curves, with the activity signal  modelled as a series of sinusoids at the rotational period of the star and its harmonics.}
{We confirm that Pr\,0211b has a nearly circular orbit ($e = 0.02 \pm 0.01$), with an improvement of a factor two with respect to the previous determination of its eccentricity, and estimate that Pr\,0211c has a mass \mplanet $\sin i = 7.9 \pm 0.2$ \mjup , a period P$>$3500 days and a very eccentric orbit (e$>$0.60). This kind of peculiar system may be typical of open clusters if the planet-planet scattering phase, which lead to the formation of HJs, is caused by stellar encounters rather than by unstable primordial orbits.
Pr\,0211 is the first multi-planet system discovered around an OC star.}{} 

\keywords{(Stars:) individual: Pr\,0211 --- Stars: planetary systems --- techniques: radial velocities, photometric}
\authorrunning{L. Malavolta }
\titlerunning{GAPS XI. - Pr\,0211 in M\,44: the first planetary system in an open cluster }
\maketitle

\section{Introduction}\label{sec:introduction}

Nearly 2\,000 extrasolar planets in 1\,300 planetary systems have been confirmed so far using a variety of techniques, prominently radial velocities (RV) and transit photometry (\url{exoplanet.org}, \citealt{Han2014}). Almost all these planets belong to isolated field stars, whose parameters (distance and age, above all) can be very uncertain. Consequently, the stellar mass and radius are affected by large errors, which directly influence the precision with which we can estimate planet parameters \citep{Casagrande2011,Santos2013}. 
Comparing dynamical models of planet formation and evolution with observations requires a good knowledge of the age of the planetary systems, which in turn can be precisely determined only for a small sample of stars using asteroseismology. The chemical signature of planet formation on the stellar host is still an open question (see, \eg , \citealt{GonzalezHernandez2013}), with the interpretation of the results relying strongly on the choice of the stellar sample and the adopted methodologies, see, for example, the opposite points of view of \cite{Adibekyan2014} and \cite{Ramirez2014}.

In principle,  searching for planets in star clusters, in particular open clusters (OCs), offers an interesting and more appropriate laboratory. OC distances and ages can be precisely determined using statistical approaches, such as isochrone fitting and gyrochronology, which is more difficult to apply to field stars \citep{Soderblom2010}.
FGK main-sequence (MS) stars are the best targets, as they have low stellar activity, negligible mass loss, and make (low mass) planet
identification easier, and planet-star interaction more straightforward to measure than in, for example, giant, evolved stars. Typically OC stars are chemically homogeneous so we can investigate the effect of the presence of planetary systems on the host star chemistry effectively. From the comparison of rotation and activity properties of planet hosts with those of other cluster members, we can get robust inference on the possible role of close-in planets on the evolution of stellar angular momentum \citep{Lanza2010b}.

Despite the many advantages offered by OCs, the number of planets discovered in OCs is still very small. 
An early RV search for planets in the Hyades cluster was attempted by \cite{Cochran2002} but the limited knowledge of the influence of stellar activity on RV measurements hampered the results.
After the discovery of two long-period massive planets around OC giant stars by \cite{Sato2007} and \cite{Lovis2007}, 
only recently new successes have been achieved
with the discovery of two hot Jupiters (HJ) around two MS stars in M\,44 \citep[hereafter Q12]{Quinn2012}, two hot Neptunes around MS stars and a long-period giant planet around a sub-giant branch star in M\,67 \citep{Brucalassi2014}, and an eccentric HJ around an MS star in the Hyades \citep{Quinn2014}. Except for the latter, eccentricities for all these planets have large uncertainties. 

In this paper, we present the first results of an observing campaign within the Global Architecture of 
Planetary Systems (GAPS, \citealt{Covino2013,Desidera2014,Poretti2015}) for the search of exoplanets around 60 stars in the open clusters M\,44 (Praesepe),  Hyades and NGC\,752. M\,44 is located at 187 parsec ($m-M=6.36$ mag), with a colour excess of $E(B-V)=0.009$ mag and an age of  $790 \pm 30$ Myr that has been estimated from isochrone fitting \citep{Brandt2015}. An independent age estimation of $578 \pm 12$ Myr has been performed by \cite{Delorme2011} using gyrochronology. 
We  focus on Pr\,0211, a $V=12.15$, G9V star hosting one of the two HJ discovered by Q12.  We  derive improved orbital parameters of Pr\,0211b,  a 1.8 \mjup\  planet with orbital period $P=2.14$ days and a RV semi-amplitude $K \simeq 300$ \ms. We  also show that the same star hosts a second, more massive planet, located at a larger distance and in a much more eccentric orbit than Pr\,0211b.

\section{Observations}\label{sec:observations}
In 2012 we started an observational campaign with HARPS-N at the Telescopio Nazionale Galileo (TNG) targeting GK stars in the open cluster M\,44, as part of GAPS. Our aims were to derive accurate and reliable eccentricities for the two known planets in the cluster, and search for additional intermediate-mass planets by pushing HARPS-N down to the limits imposed by stellar activity. 
Both goals can only be achieved with a careful observational strategy,
because of the large activity of M\,44 stars. During the observability season of the cluster, we took series of spectra on contiguous nights whenever the weather allowed us, and avoided observing the targets for just  one or two isolated nights allocated to GAPS, to properly sample the rotational period of the stars  ( $7.97 \pm 0.07$ days, \citealt{Kovacs2014}, hereafter K14).

We started observing Pr\,0211 in March 2013  and we continued observing the star until May 2015. It soon became clear that a long-term trend was present in the HARPS-N data, and therefore we further coordinated our RV campaign with TRES observations to reliably measure the RV offset between the instruments and take advantage of the earlier (2012) observations of Q12 . A nearly simultaneous photometric campaign was conducted to characterize the stellar activity. The simultaneous characterization of the eccentricity of the HJ, the quasi-periodic nature of stellar activity signal, and a possible long-period planet require an important investment of observing time spread across several years.

\subsection{Radial velocity observations. I. HARPS-N dataset}
We collected 70 HARPS-N spectra between March 2013 and May 2015. An exposure time of 1200\,s was used to ensure an average signal-to-noise ratio (\snr ) of 25 per extracted pixel at 5500 \AA, which resulted in an average radial velocity precision of 6 \ms.  We did not make use of the simultaneous Thorium-Argon calibration lamp so as not to contaminate the stellar spectra, and because the RV noise for a star of this magnitude combined with the moderate rotation of the star (\vsini $=4.8\pm 0.5$ \kms , from Q12), largely dominates the internal stability of the instrument ($\simeq 1.0$ \ms, \citealt{Cosentino2014}). 
As a measure of precaution, we used tight constraints for the Moon position and the sky background illumination. The target was not observed when the distance from the Moon was less than 60$^{\circ}$ or when the sky was brighter than V$=19~ \textrm{mag}\ \textrm{arcsec}^{-2}$. 

Data reduction and RV extraction were performed using the most recent HARPS-N Data Reduction Software. We used the standard G2 mask to determine the weighted cross-correlation function (CCF, \citealt{Baranne1996,Pepe2002}). We used the CCF noise provided by the pipeline as estimate of the RV uncertainty. 

\subsection{Radial velocity observations. II. TRES dataset}

The dataset included here is a continuation of the one presented in
Q12, obtained with the Tillinghast Reflector Echelle Spectrograph (TRES, \citealt{Furesz2008}) between January and April 2012, one year before the HARPS-N observations. We refer to Q12 for a detailed description of the instrument and the data reduction details. In addition to the 18 RVs of the discovery paper, 36 new RVs with an average error of $\simeq 20$ \ms\ were gathered during the same temporal window of HARPS-N observations. TRES observations cover a time span of $\simeq 1200$ days \ie , 3.3 years.

\subsection{Photometry}
Concurrently with the RV campaign we performed a photometric follow-up to determine the rotational period of the star and monitor its long-term activity level. Observations were gathered with the \textit{Wide Field STELLA Imaging Photometer} (WiFSIP) mounted on \textit{STELLar Activity-I} telescope at Teide Observatory, in Tenerife \citep{Strassmeier2004,Weber2012}.
We obtained data spread over two months for every year between 2013 and 2015, with measurements in the Bessel $B$ band over four to six separate pointings for a total of  56 (2013), 26 (2014), and 31 (2015) epochs. During 2015 the star was observed in the Sloan $r$ band as well.

Data reduction, aperture photometry, and correction of light curves for systematics were performed following \cite{Nascimbeni2014}. We used the generalized Lomb-Scargle periodogram (GLS, \citealt{Zechmeister2009}) to verify that the rotational period of $\simeq 8$ days was visible in our data.

\section{Stellar parameters}\label{sec:stellar_parameters}
Atmospheric parameters have been determined using the classical equivalent width (EW) approach on the HARPS-N coadded spectrum with \snr\ $\simeq 320$.
EWs were measured using the latest version of ARES \citep{Sousa2015} in automatic mode. Atmospheric parameters were determined using the 2014 version of the local thermodynamic equilibrium code MOOG\footnote{Available at \url{http://www.as.utexas.edu/~chris/moog.html}} \citep{Sneden1973} and the Kurucz model atmospheric grid\footnote{Available at \url{http://kurucz.harvard.edu/grids.html}} with the new opacity distribution function (ODFNEW, \citealt{Castelli2004,Kurucz1992}). We used the iron linelist from \cite{Sousa2011}, but we modified the values of the atomic line oscillator strength  \loggf\ to match the solar EWs with the elemental abundances of \cite{Asplund2009} (Table~\ref{table:linelist}).

\onltab{
\begin{table}
\caption{Linelist with the oscillator strength \loggf\ updated to reflect the new solar abundances in the 2014 version of MOOG. All the values except the \loggf\ are taken from \url{http://www.astro.up.pt/~sousasag/ares/}. 
The full table is available on-line, a portion is shown here for reference.}              
\label{table:linelist}      
\centering                                      
\begin{tabular}{c c c c c c}          
\hline\hline                        
\noalign{\smallskip}
\multicolumn{2}{c}{Element} & $\lambda \,[\AA ]$ & $\chi \, [$eV$]$ & \loggf  & EW $[$m$\AA ]$\\
\noalign{\smallskip}
\hline                                 
\noalign{\smallskip}
    FeII & 26.1 &  4508.28 &     2.86  &   -2.539  &   87.3 \\
    FeII & 26.1 &  4520.22 &     2.81  &   -2.688  &   81.9 \\
    FeI  & 26.0 &  4523.40 &     3.65  &   -1.958  &   44.2 \\
     ... & ... & ... & ...& ... & ... \\
\noalign{\smallskip}
\hline
\end{tabular}
\end{table}
}

Effective temperature \teff\ and microturbulent velocity $\xi$ were determined by minimizing the trend of iron abundances from individual lines with respect to their excitation potential and reduced EW respectively, while the surface gravity \logg\ was derived by imposing the same abundances for neutral and ionized iron lines. The final parameters with their errors were determined iteratively as in \cite{Dumusque2014}.
We checked the reliability of the automatic EW measurement by determining the photospheric parameters with the EWs extracted for a range of the \texttt{rejt} parameter, and verifying that the value automatically calculated by ARES was in fact providing the photospheric parameters with the lowest error. 
We found \teff\ $=5270 \pm 60$ K, \logg\ $=4.46 \pm 0.10$, $\xi=1.1 \pm 0.1$ \kms\ and \gfeh\ $=0.18 \pm 0.04$, where the quoted uncertainties include possible systematic errors associated with this method \citep{Sousa2008}. These values are consistent within the errors with the ones listed by Q12 and obtained using a synthetic match approach (SPC, \citealt{Buchhave2012}), hence we used the weighted average of the two determinations as our final set of photospheric parameters in Table~\ref{table:stellar_parameters}.
We determined the projected rotational velocity \vsini\ by fitting synthetic spectra of seven isolated iron lines on the HARPS-N coadded spectrum of Pr\,0211. We assumed the atmospheric parameters quoted in Table~\ref{table:stellar_parameters}, a spectrograph resolution of $R=115000$, a macroturbulent velocity \vmacro $ = 2.3 \pm 0.7$ \kms\ from the calibration of \cite{Doyle2014}, and a limb darkening coefficient of 0.65 \citep{Gray2005book}. We obtain \vsini $= 5.1 \pm 0.3$ \kms , where the uncertainty includes the error contribution from \vmacro .

The mass and radius of the star were derived using four sets of isochrones to analyse systematic differences between stellar models: PARSEC \citep{Bressan2012}, Dartmouth \citep{Dotter2008}, Yonsei-Yale \citep{Yi2001} and BaStI \citep{Pietrinferni2004}. We followed a Monte Carlo (MC) approach to take the uncertainties in temperature and metallicity into account. We took the average of the values as final parameters, while the associated errors are obtained by adding in quadrature the sample standard deviation of the measurements and their average error. We obtained a stellar mass of \mstar\ $=0.935 \pm 0.013$ \msun\  and a radius of  \rstar\ $=0.827 \pm 0.012$ \rsun .

We performed a second determination with the same approach using $B-V=0.87 \pm 0.01$, which is an average of the values obtained by \cite{Johnson1952} 
and \cite{Upgren1979} with a conservative choice for the associated errors, and the value for $E(B-V)$ quoted in Section~\ref{sec:introduction}. In this case we obtained \mstar\ $=0.933 \pm 0.017$ \msun\ and \rstar\ $=0.825 \pm 0.016$ \rsun , which was perfectly consistent with the spectroscopy-based value. Assuming an age of $\simeq 800$ Myr instead of $\simeq 600$ Myr caused a decrease of 0.002 \msun\ in the derived masses and an increase of 0.002 \rsun\ for the stellar radius, \ie , the error on the age is negligible for the mass determination.
In the following, we used the spectroscopy-based value for the mass and radius of the star.

We took advantage of the knowledge of the photometric period \prot\ from K14 and the projected rotational velocity \vsini\ to estimate the value of the inclination of the stellar spin axis \istar\ to the line of sight. We included all the source of uncertainties through an MC approach, although we verified that the \vsini\ is the biggest contributor to our error budget.

\begin{table}
\caption{Parameters of the star.}              
\label{table:stellar_parameters}      
\centering                                      
\begin{tabular}{l c c}          
\hline\hline                        
\noalign{\smallskip}
Parameter & Value & Unit\\
\noalign{\smallskip}
\hline                                 
\noalign{\smallskip}
\teff\   & $5300 \pm 30$   & K\\
\logg\   & $4.51 \pm 0.05$ & dex \\
\vmicro\ & $1.1 \pm 0.1$   & \kms \\
\gfeh\   & $0.18 \pm 0.02$ & dex \\
\vsini\  & $5.1 \pm 0.3$   & \kms \\
\mstar\  & $0.935 \pm 0.013$ & \msun  \\  
\rstar\  & $0.827 \pm 0.012$ & \rsun \\ 
Age      & $578  \pm 12$\tablefootmark{a}  & Myr \\ 
         & $790  \pm 30$\tablefootmark{b}  & Myr \\ 
\istar\  & $76 \pm 11 $ & deg \\
\logRHK\  & $ -4.36 \pm 0.04 $ & dex \\ 
\noalign{\smallskip}
\hline
\end{tabular}
\tablebib{
\tablefoottext{a}{\citet{Delorme2011}.}
\tablefoottext{b}{\citet{Brandt2015}.}
}

\end{table}
\section{Analysis of RVs}\label{sec:rv_analysis}
\begin{figure}[htbp]
\includegraphics[width=\linewidth]{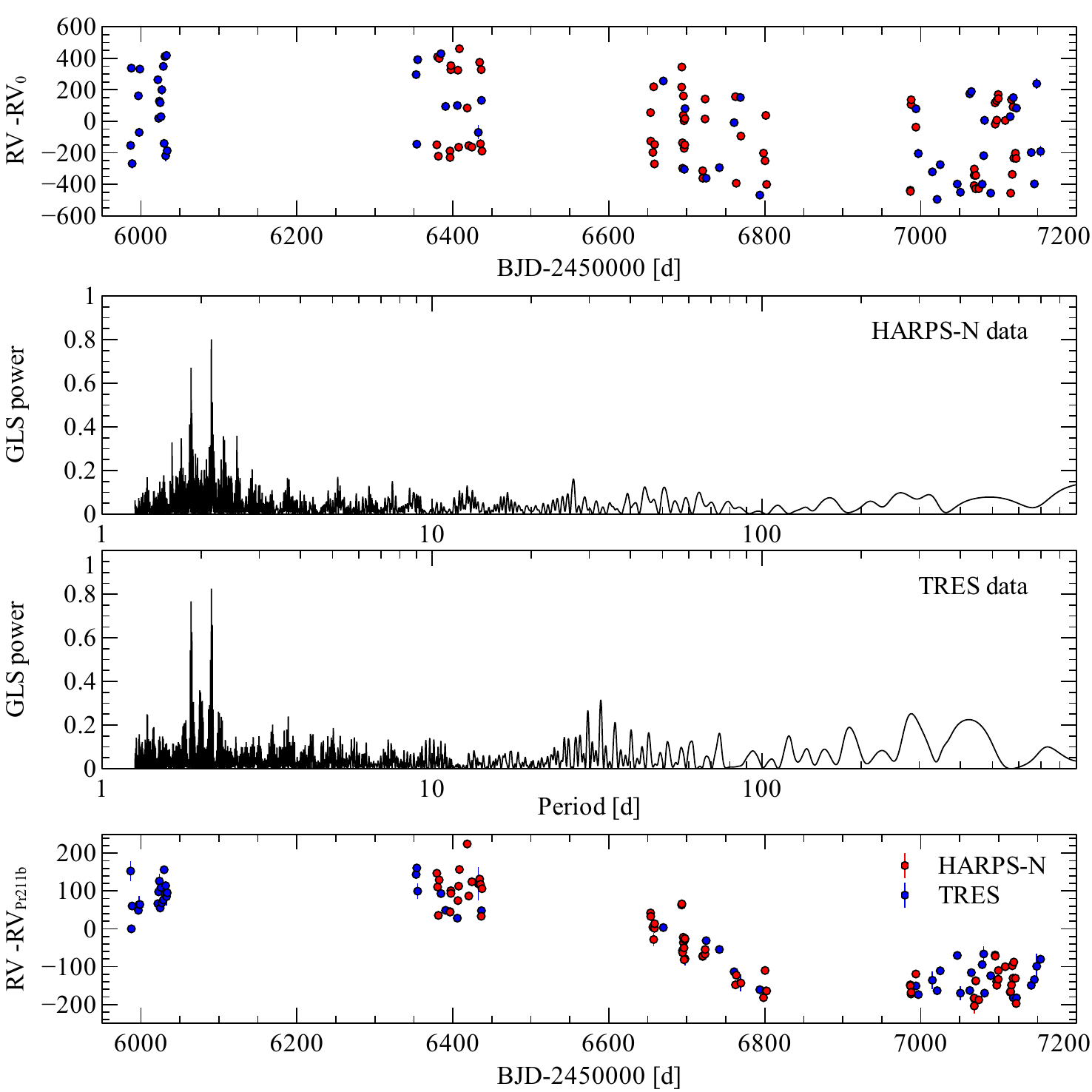}
\caption{ Top panel: RV data obtained with HARPS-N (red dots) and TRES (blue dots) as a function of time. Middle panels: GLS periodogram of the observed RVs (black lines) for the HARPS-N and TRES datasets, respectively. Lower panel: residuals of the RVs after removing the signal from Pr\,0211b using the solution found by GLS.}
\label{fig:GLS_RV_periods}
\end{figure}

The radial velocity data we collected from HARPS-N  and TRES are shown in the upper panel of Figure~\ref{fig:GLS_RV_periods}, after removing the instrumental offset between the two datasets, as described in Section~\ref{sec:orbit_fitting}. The RVs, along with the activity indexes and photometry, are available online in the format shown in Table~\ref{table:RV_table}, \ref{table:asymmetry_table}, and \ref{table:activity_table}.
Before performing a full orbital fit we  analysed the HARPS-N and TRES data independently with the GLS algorithm (second and third panel of Figure~\ref{fig:GLS_RV_periods} respectively). 
The periodograms of both datasets confirm the presence of the planet with period of 2.14 days and RV semi-amplitude of $\simeq 300$ \ms\ , as discovered by Q12. We also note that the alias at 1.86 days supplies  much poorer folded RV curves.

The residuals obtained after subtracting a sinusoid at 2.14 days reveal the presence of a trend with the same amplitude in both datasets. The deviation of the trend from a sinusoidal shape makes the period estimate from GLS highly unreliable and a Keplerian curve should be used instead. However, before proceeding to an orbital fit, we need to assess if the long-period trend may be caused by stellar activity.

\onltab{
\begin{table}
\caption{RV data collected with HARPS-N (a) and TRES (b). Instrumental offsets have not been applied. The full table is available on-line, a portion is shown here for reference.}              
\label{table:RV_table}      
\centering                                      
\begin{tabular}{c c c c}          
\hline\hline                        
\noalign{\smallskip}
BJD$_{\rm UTC}$ & RV & $\sigma$ RV  & Instrument  \\     
 $[$d$]$  &  $[$\ms $]$  &  $[$\ms $]$  &  \\
\noalign{\smallskip}
\hline                                 
\noalign{\smallskip}
    2456379.42879  &   34880.0     &    5.2  & a \\ 
    2456380.43289  &   35438.1     &    8.6  & a \\
    2456381.39909  &   34807.5     &    6.4  & a \\
     ... & ... & ... & ... \\
\noalign{\smallskip}
\hline
\end{tabular}
\end{table}
}

\onltab{
\begin{table*}
\caption{CCF Asymmetry indicators extracted from HARPS-N spectra. The errors associated to  BIS, BIS$-$, BIS$+$ , $\Delta$V  and   $V_{\rm span}$ are twice  $\sigma_{\rm RV}$, so they have not been reported in the table. The full table is available online, only a portion is shown for references.}
\label{table:asymmetry_table}      
\centering                                      
\begin{tabular}{c c c c c c c c c c c c }     
\hline\hline                        
\noalign{\smallskip}
 BJD-2450000.0 & RV$_{\rm res}$ & $\sigma_{\rm RV}$ & BIS & BIS$-$ & BIS$+$ & $\Delta$V &  $V_{\rm span}$ & FWHM & $\sigma_{\rm FWHM}$ & $V_{\rm asy} \cdot 10^3$ &  $\sigma_{V_{\rm asy}} \cdot 10^3$ \\
\noalign{\smallskip}
\hline                                 
\noalign{\smallskip}
     6379.428774  &  151.14 &  5.2  &  $-$24.6  &  $-$13.3 &   $-$40.9  &   43.1 &  $-$26.0  &  9.403 &  0.010   &  39.15 &   0.37 \\
     6380.432878  &   95.54 &  8.6  &   81.4  &   52.6 &    86.2  &  $-$83.8 &   66.4  &  9.343 &  0.017   &  39.12 &   0.51 \\
     6381.399079  &   69.21 &  6.4  &   $-$3.0  &    2.3 &    $-$5.4  &    1.2 &    5.2  &  9.299 &  0.012   &  38.54 &   0.44 \\
... & ... &  ... &  ... &  ... &  ... &  ... &  ... &  ... &  ... &  ... &  ... \\
\noalign{\smallskip}
\hline
\end{tabular}
\end{table*}
}

\onltab{
\begin{table*}
\caption{Activity indicators extracted from HARPS-N spectra. The full table is available online, only a portion is shown for references.}
\label{table:activity_table}      
\centering                                      
\begin{tabular}{c c c c c c c c c c c c }     
\hline\hline                        
\noalign{\smallskip}
 BJD-2450000.0 & \logRHK & $\sigma_{\log {\rm R}^{\prime}_{\rm HK} }$ & H$\alpha$ & $\sigma_{{\rm H}\alpha}$ & HeI & $\sigma_{\rm HeI}$ & NaID & $\sigma_{\rm NaID}$ \\ 
\noalign{\smallskip}
\hline                                 
\noalign{\smallskip}
6379.428774  & $-$4.351 &  0.013  &    0.2462  &   0.0022  &    0.489 &     0.008   &         -     &      -    \\
6380.432878  & $-$4.273 &  0.023  &    0.2412  &   0.0033  &    0.489 &     0.013   &        0.2123 &    0.0050 \\
6381.399079  & $-$4.377 &  0.021  &    0.2369  &   0.0029  &    0.482 &     0.011   &        0.2000 &    0.0039 \\
... & ... &  ... &  ... &  ... &  ... &  ... &  ... &  ...  \\
\noalign{\smallskip}
\hline
\end{tabular}
\end{table*}
}

\section{Asymmetry and activity indexes}\label{sec:asymmetry_activity}
Line profile variation indicators were derived from the CCF using a variety of methods.
The full width at half maximum (FWHM) of the CCF, the contrast of the CCF, and the bisector inverse slope (BIS) were determined automatically by the HARPS-N pipeline \citep{Cosentino2012}. The velocity span $V_{\rm span}$ \citep{Boisse2011}, the BIS$^+$ and BIS$^-$ \citep{Figueira2013}, and $\Delta V$ \citep{Nardetto2006} are derived using the scripts provided by \cite{Santos2014}. We associated as conservative errors twice the CCF$_{\textrm{noise}}$ for all these indicators. 
Additionally we used a modified version of the $V_{\rm asy}$ activity indicator (Lanza \etal\ in prep.\footnote{The algorithm was introduced at the Extreme Precision Radial Velocity workshop and a description is available at \url{https://sites.google.com/a/yale.edu/eprv-posters/home}}), which avoids the explicit presence of the RV in its definition, in contrast with the original formulation of \cite{Figueira2013}.

In the analysis, we included activity indexes that are directly derived from the observed spectrum, rather than the CCF, and which are specifically designed to monitor the magnetic and chromospheric activity of the star. The \logRHK\ index \citep{Noyes1984} was determined using a script that came with the HARPS-N DRS \citep{Lovis2011}, using the $(B-V)_0$ color index quoted in Section~\ref{sec:stellar_parameters}.
The \halpha\ index \citep{GomesDaSilva2011,Robertson2013}, the NaID index \citep{Diaz2007} and the HeI index \citep{Boisse2009} were measured using our own implementation of the algorithms.

\begin{figure}[htbp]
\includegraphics[width=\linewidth]{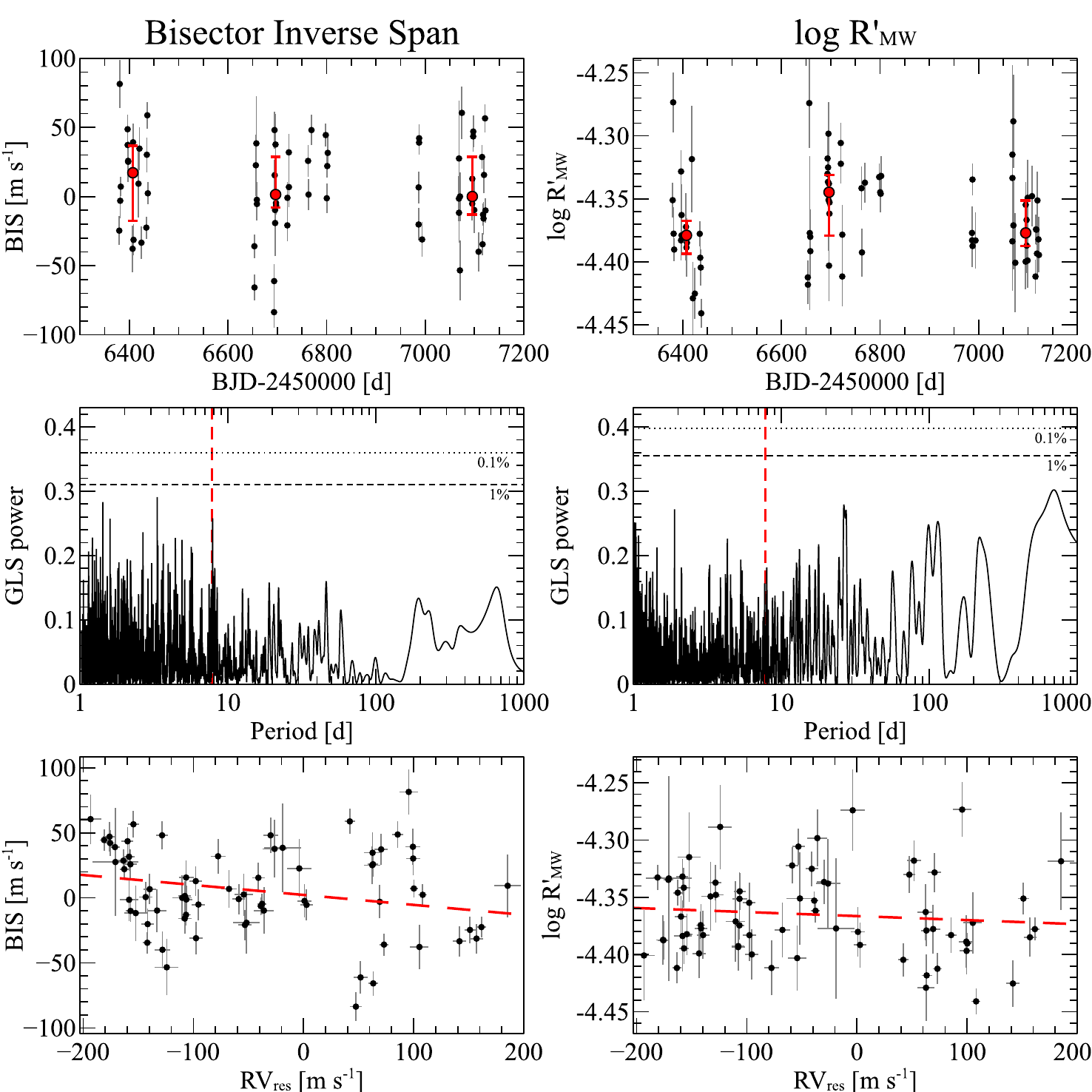}
\caption{Analysis of the bisector inverse span (panels on the left side) and the \logRHK\ index (panels on the right side). Upper panels: the indicators as a function of time, the seasonal medians with the first and third quartiles indicated in red. Middle panels: GLS periodograms of the indexes, the rotational period of the star is indicated with a red vertical line. The 1\% and 0.1\% FAP levels are displayed as dashed and dotted horizontal lines, respectively. Lower panels: indicators as a function of RV, after removing the signal of Pr\,0211b. The best fit is represented by the dashed red line}
\label{fig:asy_act_index}
\end{figure}

In Figure~\ref{fig:asy_act_index} we included only the BIS and \logRHK\ as representative of the other indexes listed above since they all show similar behavior. In the upper panels we plot the activity indexes as a function of time to check if there is any correlation with the long-period RV variation observed in the data (see the lower panels in  Figure~\ref{fig:GLS_RV_periods} for comparison). 
In the middle panels the GLS periodograms of the corresponding index are computed. The 1\% and 0.1\% false alarm probability (FAP) were estimated with a bootstrap approach.

In the lower panels of Figure~\ref{fig:asy_act_index} we investigated the presence of correlations between the asymmetry and activity indexes with RVs, after removing the signal of Pr\,0211b.
We determined the Spearman's rank correlation coefficient $\rho$, the slope of the linear fit $m$ with its error and the $p$-value using the weighted least-square regression that was implemented in the {\tt StatsModels} package\footnote{Available at \url{http://statsmodels.sourceforge.net/}}. 
Despite the signature of activity added to the RVs on the timescale of the rotational period, in all cases the correlation rank is close to zero, and the slope of the best fit is consistent with the lack of any trend within one or two $\sigma$, depending on the index that is  considered (Table~\ref{table:rho_m} and Figure~\ref{fig:actindex_RVs}).

\onltab{
\begin{table}
\caption{Spearman's correlation rank, slope of the linear fit and its $p$-value for several asymmetry and activity indexes as a function of RVs, after he signal of the first planet has been removed.}
\label{table:rho_m}      
\centering                                      
\begin{tabular}{l c c c }          
\hline\hline                        
\noalign{\smallskip}
Index & $\rho$ & $m$ & $p$-value \\     
\noalign{\smallskip}
\hline                                 
\noalign{\smallskip}
BIS           & $-0.23$ & $-0.08 \pm 0.04$   &  0.11 \\
RV span       & $-0.21$ & $-0.05 \pm 0.03$   &  0.17 \\
$\Delta V$    & $-0.18$ & $-0.001 \pm 0.001$ &  0.70 \\ 
V$_{\rm asy}$ & $ 0.17$ & $0.0004 \pm 0.001$ &  0.31 \\
\logRHK       & $-0.10$ & $(-4 \pm 4) \cdot 10^{-5}$ & 0.39 \\  
\halpha       & $-0.03$ & $(-7 \pm 5) \cdot 10^{-6}$ & 0.57 \\
HeI           & $ 0.16$ & $( 6 \pm 4) \cdot 10^{-6}$ & 0.19 \\
NaID          & $-0.06$ & $(-2 \pm 4) \cdot 10^{-6}$ & 0.67 \\
\noalign{\smallskip}
\hline 
\end{tabular}
\end{table}
}

\onlfig{
\begin{figure}[htbp]
\includegraphics[width=\linewidth]{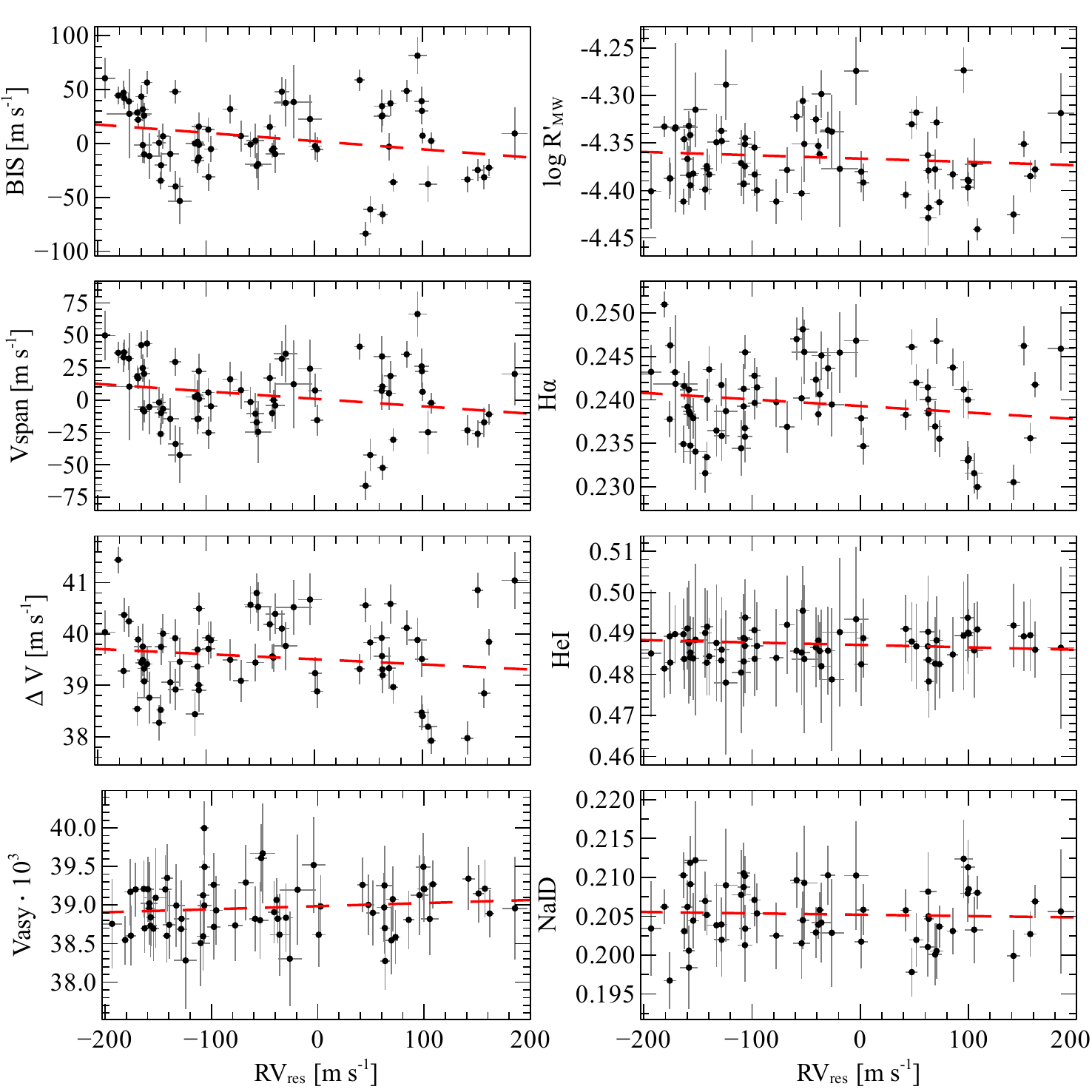}
\caption{CCF asymmetry indexes (panels on the left side) and chromospheric activity indicators (panels on the right side) as a function of RVs, after removing the signal of Pr\,0211b. The best fit is represented by the dashed red line.}
\label{fig:actindex_RVs}
\end{figure}
}

We assume here that any physical effect, whether spots or flares or magnetic cycles, capable of producing a peak-to-peak variation of $\simeq 280$ \ms\ in the RVs, should have left its imprint on at least one of the several indicators we have at our disposal. However, in all cases we can see that all the indicators are dominated by short-term  variations, the effect of which has been averaged out by our observational strategy, see Section \ref{sec:observations}. While peaks at $\simeq 700$ days can be seen in the $\Delta V$, \logRHK\ and \halpha\ indexes, their significance is below the 1\% threshold and they are not correlated to the longer period variation that is observed in the RVs (Figure~\ref{fig:asyindex} and \ref{fig:actindex}).
We note that the signal associated with the rotational period is visible in the periodograms, although not highly significant. This is, however, expected, given the low \snr\ of our measurements, combined with the slight loss of coherence of the short-period activity signal across several observing seasons, as already observed in \cite{Santos2014}.

\onlfig{
\begin{figure}[htbp]
\includegraphics[width=\linewidth]{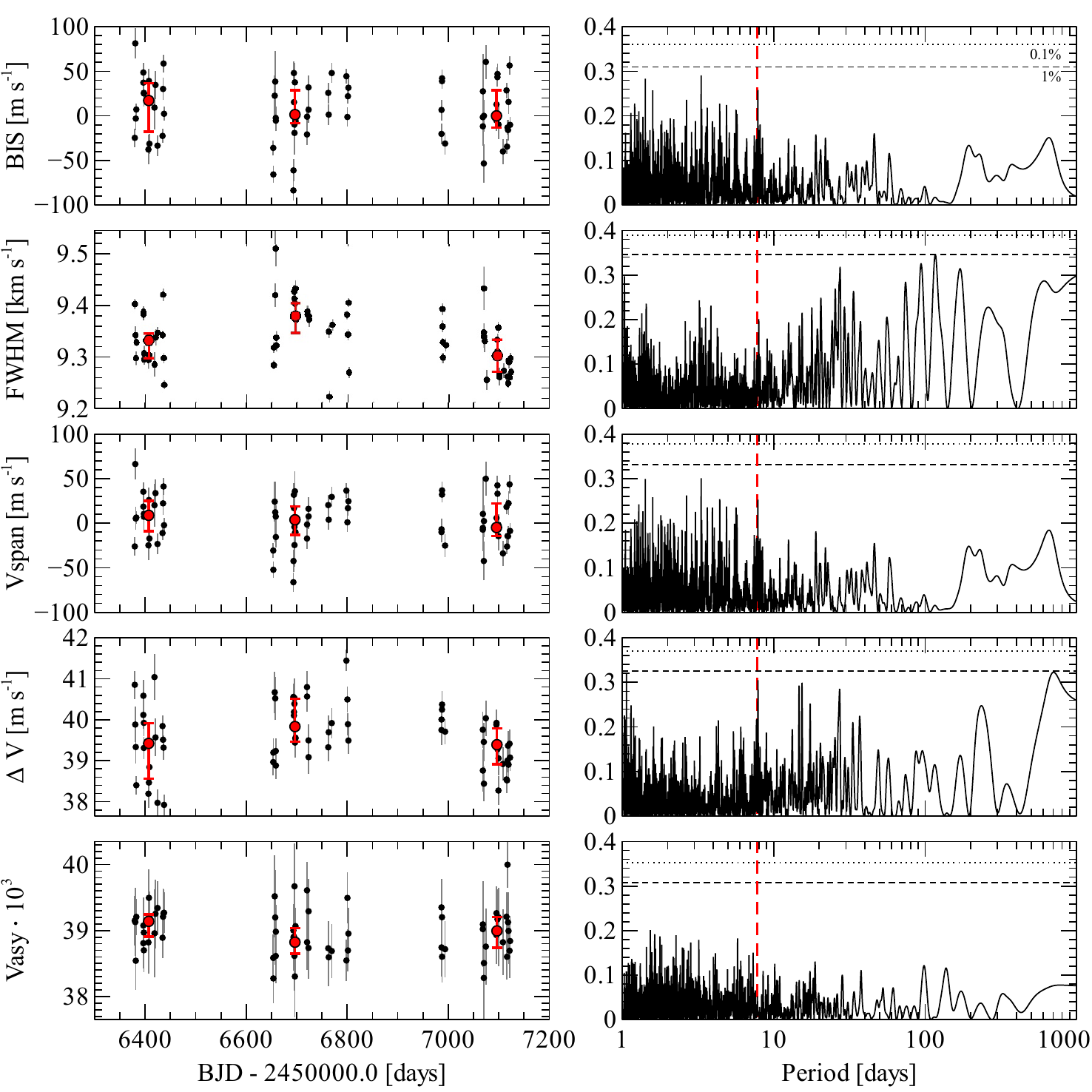}
\caption{Asymmetry indicators as a function of time (panels on the left) and their corresponding GLS periodogram (panels on the right). From top to bottom the bisector inverse span BIS, the FWHM of the CCF, the RV span, the $\Delta V$ and V$_{\rm asy}$ are shown. Seasonal medians with the first and third quartiles are  indicated in red. The rotational period of the star is indicated with a red vertical line. The 1\% and 0.1\% FAP levels are displayed as dashed and dotted horizontal lines, respectively. The extra power in the FWHM periodogram is due to a drift in the instrumental resolution, corrected in March 2014 (JD $\simeq 2456370$).}
\label{fig:asyindex}
\end{figure}
}

\onlfig{
\begin{figure}[htbp]
\includegraphics[width=\linewidth]{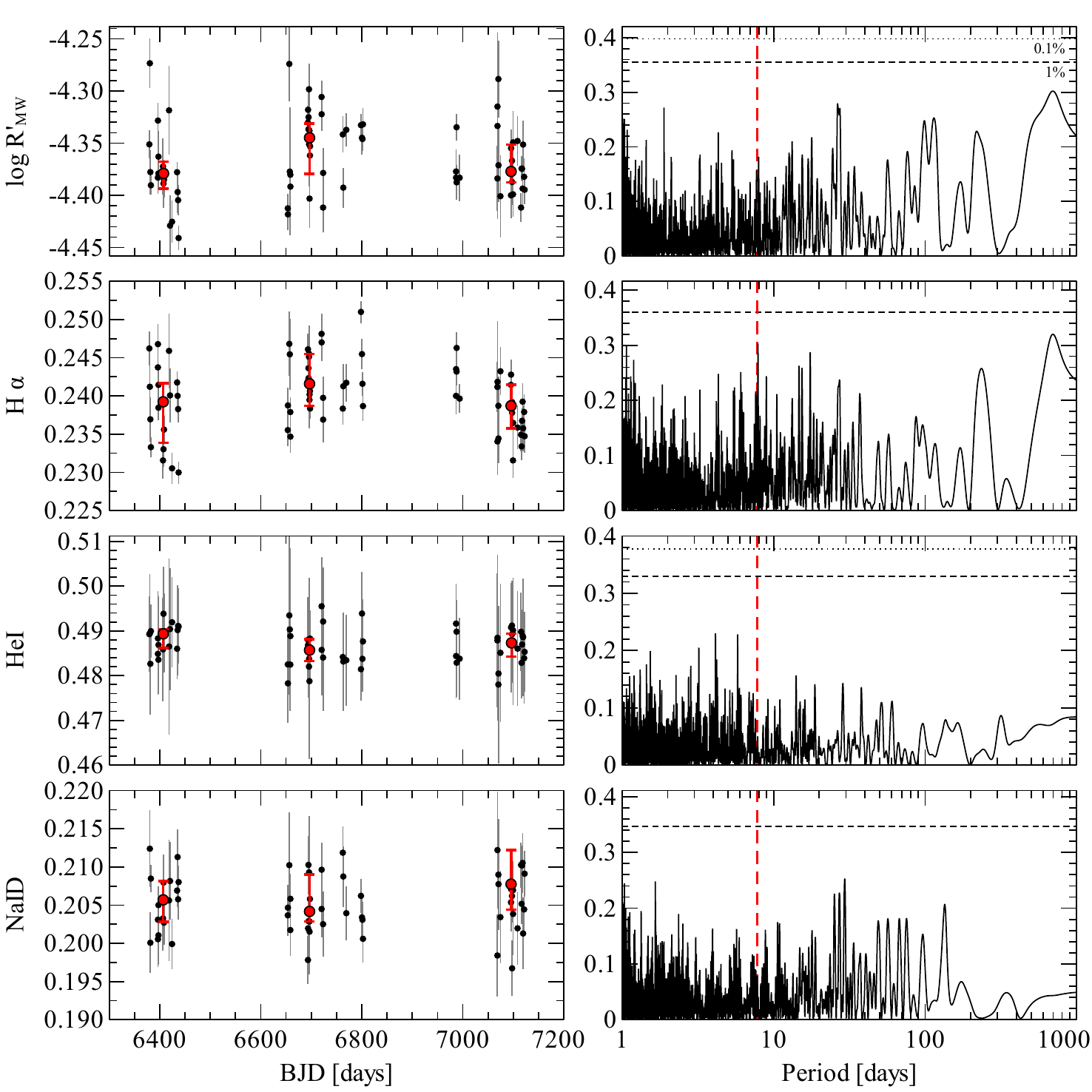}
\caption{{\footnotesize As in Figure~\ref{fig:asyindex}, but with the activity indicators \logRHK , \halpha , HeI and NaID (from top to bottom).}}
\label{fig:actindex}
\end{figure}
}

The marginal slope in the activity indexes advocates against a stellar origin for the long-term RV variation trend. While long-term trends in the chromospheric activity have been observed in several stars, their effect on RVs is usually well within 100 \ms\ and correlations between RV and activity indexes are seen even for moderately active stars \citep{Santos2000}.
Despite the high level of activity of the star, the presence of a planet remains the simplest explanation to the long-term RV variation that we observed.  

\section{A two-planet system revealed}\label{sec:orbit_fitting}
We performed the combined TRES and HARPS-N RV fit with the affine-invariant Markov Chain Monte Carlo (MCMC) ensemble sampler \texttt{emcee} code (\citealt{ForemanMackey2013}, hereafter FM13), coupled with the differential evolution genetic algorithm \textrm{PyDE}\footnote{Available at \url{https://github.com/hpparvi/PyDE}} to determine the global optimum solution  as the initial point of the MCMC chains.

We take instrumental systematics into account by including an offset term between dataset, and adding in quadrature a jitter term to the errors that are associated with the measurements. Different jitter and offset parameters are used for datasets that come from different instruments, even if they are measuring the same quantity, and set as free parameters during the fitting process. We assumed the HARPS-N offset as the systemic velocity of the star $\gamma$.

We first analysed our data with a two-planet model that does not include any treatment for activity and works exclusively with the RVs that were obtained with TRES and HARPS-N. Our model has a total of 14 parameters, which are the Keplerian orbital parameters for two planets plus the systemic velocity of the star, a RV jitter term for each dataset, and finally the difference in the RV zero-point between HARPS-N and TRES data. 
We followed \cite{Eastman2013} by fitting the period $P$ and the semi-amplitude $K$ of the signal in the logarithmic space, and determining the eccentricity $e$ and the argument of pericenter $\omega$ by fitting $\sqrt{e} \cos{\omega} $ and $\sqrt{e} \sin{\omega}$. 
Instead of using the time of periastron $T_{\rm peri}$, which has to be constrained within one orbital period to facilitate chains convergence, we fitted the phase of the orbit $\phi = M_0 + \omega$, where $M_0$ is the mean anomaly with respect to an arbitrary reference time $T_{\rm ref}$.
 
Convergence of the chains is considered as achieved when the Gelman-Rubin statistic $\hat{R}$ is lower than 1.03 \citep{Gelman1992,Ford2006}. Additionally we verified that the acceptance fraction for all the walkers was in the range $[0.20,0.50]$, as suggested by FM13, as well as visual inspection of the individual chains. 
We discarded the section of the chains where the convergence criteria was not yet satisfied as burn-in phase.
The median values and the standard deviation of the posterior distributions have been determined after applying a thinning factor of $\simeq 100$ to the chains, \ie\ , the average autocorrelation time. We used the 34.13 percentile at each side of the median as an approximation of the standard deviation.

\begin{figure}[htbp]
\includegraphics[width=\linewidth]{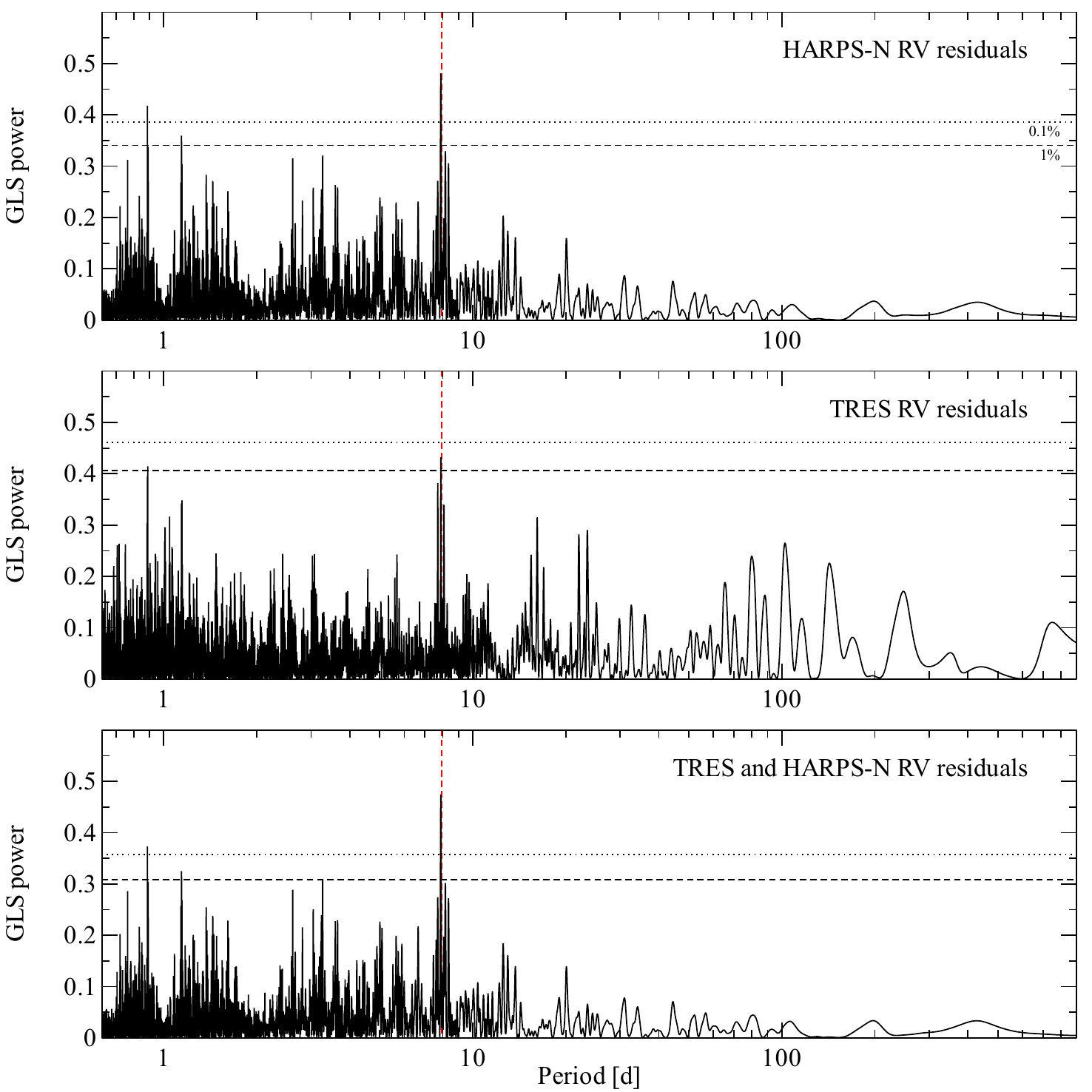}
\caption{{\footnotesize GLS periodograms of the RV residuals after subtracting the 2-planet solution, for HARPS-N data (upper panel), TRES data (middle panel) and the two datasets combined (lower panel). The rotational period of the star is indicated by a dotted vertical line.}}
\label{fig:rv_residuals_GLS_k0}
\end{figure}

The results are listed in Table~\ref{table:orbit_result}. For Pr\,0211b, we determined a projected mass of $1.88 \pm 0.03$ \mjup\, and an eccentricity that is consistent with zero. If we assume that the orbit of the planet and the spin of the star are aligned, as is expected for HJs around  a star of \teff $\simeq 5300 $K \citep{Dawson2014}, using our previous determination of \istar\ we obtain a true mass for the planet between $1.85$ and $2.05$ \mjup , well within the planetary mass range.
For Pr\,0211c, we determine a minimum mass of $7.79 \pm 0.33$ \mjup , a very high eccentricity ($e > 0.6$) and an orbital period in the range between $3000$ and $9400$ days.
The effect of the activity can be envisaged in the large jitter term ($\simeq 30$ \ms ) compared to the internal errors. This value is in agreement with the estimate provided by \cite{Santos2000} for the observed \logRHK . Activity is the cause of the prominent peak at $7.9$ days in the periodograms of both HARPS-N and TRES RV residuals (Figure~\ref{fig:rv_residuals_GLS_k0}).

\begin{table}
\caption{Orbital parameters of the two planets, obtained by fitting
  two Keplerian orbits and no signal for the activity. Best-fitting parameters are shown as fitted, additional parameters that were computed starting from the fitted parameters are tagged as derived.}
\label{table:orbit_result}      
\centering                                      
\begin{tabular}{l c c c }          
\hline\hline                        
\noalign{\smallskip}
Parameter & Pr\,0211b & Pr\,0211c & Note  \\     
\noalign{\smallskip}
\hline                                 
\noalign{\smallskip}
 P $[\textrm{days}]$        & $2.14610 \pm 3 \cdot 10^{-5}$ & $4850^{+4560}_{-1750}$ & (a)  \\   
 K $[$\ms$]$                & $309.7 \pm 4.2$ & $138 \pm 7$ & (a)  \\
 $\phi [deg]$               & $206.0  \pm 0.8 $       &  $106.7 \pm 6.7$ & (a)  \\
 $\sqrt{e} \sin{\omega}$    & $0.016 \pm 0.083$          & $0.78 \pm 0.09$ & (a)  \\
 $\sqrt{e} \cos{\omega}$    & $0.031 ^{+0.078}_{-0.086} $          &  $-0.30 \pm 0.11$ &(a)  \\
 e                         & $0.011^{+0.012}_{-0.008}$   & $0.71 \pm 0.11$ & (b) \\
 $\omega$ $[\textrm{deg}]$ & $17^{+87}_{-111} $         & $111 \pm 9 $ & (b) \\
 \mplanet $\sin i ~[$\mjup $]$  & $1.88 \pm 0.03$  & $7.79 \pm 0.33 $& (b)\\              
 $a $ $[$AU$]$                  &  $0.03176 \pm 0.00015$  &  $5.5_{-1.4}^{+3.0}$ & (b)\\ 
$T_{\rm peri}$ $[$d$]$             & $2456678.8 \pm 0.5$ & $2456736 \pm 22 $ & (b)\\ 
\noalign{\smallskip}
\hline
\noalign{\smallskip}
Parameter & HARPS-N & TRES & kind \\
\noalign{\smallskip}
\hline
\noalign{\smallskip}
$\gamma$ $[$\ms $]$  & $35029 \pm 10 $ & $135 \pm 9$ & (a) \\
RV$_{\rm jitter}$  $[$\ms $]$ & $33 \pm 3 $ & $ 26 \pm 5 $ & (a) \\
 $T_{\rm ref}$ & \multicolumn{2}{c}{2456679.97345}     & (c)  \\    
\noalign{\smallskip}
\hline
\end{tabular}
        \tablefoot{
         (a) Fitted
         (b) derived 
         (c) fixed.
         } 
\end{table}

\section{Simultaneous activity and orbital fit}\label{sec:activity_orbit}
In the previous section we included a RV jitter term to model both instrumental errors  and  stellar activity, thus implicitly assuming that the latter behaves as white noise. We know, however, that this assumption does not hold because of the correlated noise that was induced by activity variations, as seen in photometry. In principle a better precision of the orbital parameters can be obtained if the activity is properly modeled.
To do so, we developed a code\footnote{Available at \url{https://github.com/LucaMalavolta/PyORBIT}} that can simultaneously handle datasets of different kinds, such as radial velocity, activity indexes, and photometry, as well as datasets that measure the same quantity but from different instruments, \eg , RVs obtained with two spectrographs or photometry obtained in several bands.

To include the effect of activity in our analysis, we followed the approach of \cite{Boisse2011}. In this approach, the RV activity is modelled as the sum of sinusoidal signals at the rotational period \prot\ of the star and its first harmonics.The rotational period and the phases of the sinusoids are constrained by the photometry, while the amplitude of the signal in RVs is left as a free parameter. The orbits of the planets and the activity signals are fitted simultaneously with the MCMC approach described in Section~\ref{sec:orbit_fitting}. In addition to the parameters previously introduced, we must include an offset term for each observational season in each photometric band, and two different jitters because of a substantial technical improvement of WiFSIP  between the 2013 and 2014 observing seasons.

The rotational period and the phase of each sinusoid are common to all datasets, but we must take into account a possible delay on the effect of activity on the stellar line shape and the brightness of the star \citep{Queloz2001}. We simulated several spot configurations  with SOAP 2.0 \citep{Dumusque2014b}, using the astrophysical parameters of our star and the epochs of our datasets to sample the simulations, and verified that our activity model was able to reproduce the simulated data points.

The amplitudes of the sinusoids for HARPS-N data and TRES data are treated as independent parameters, since activity is a color-dependent effect and the two instruments are using different spectral ranges to determine the RVs. For the same reason, the amplitudes for photometric $B$ and $r$ bands are treated separately. Each season covered by photometric observations was fitted independently. 
Owing to the lack of simultaneous photometry, we excluded TRES data gathered in 2012 from the activity modeling and a separate RV jitter was introduced.
Finally, for \prot\ , we decided to use one free parameter for the whole temporal span since the periods for each observing season were consistent at 1-$\sigma$ within each other, when left as free parameters in the same fit. Each season is still characterized by independent values for the phases and amplitudes of the sinusoids. We assumed that the phases of the rotational signals are constant within an observational season, which is not strictly true, since these parameters are expected to change smoothly with time while spots/flares move on the surface of the star. We performed a preliminary fit using two harmonics of \prot , then we excluded those harmonics which amplitudes were consistent with zero. We finally decided to use one harmonic for the HARPS-N dataset and just \prot\ for the TRES RVs.
In the fit we did not include any activity index, since no clear signal is visible at \prot\ and its harmonics.
As such, we have a total of 35 additional parameters.
Results of the analysis are reported in Table~\ref{table:orbit_act_result}.

\begin{table}
\caption{Orbital parameters of the two planets, obtained by including activity in the global fit. $K$ denotes the semi-amplitude of the RV sinusoid at the given harmonic of the stellar orbital period, $A$ the semi-amplitude of the photometric sinusoids, $\phi$ the phases of the sinusoids (in common between RV and photometry). Only one parameter has been used for the jitter term in the $B$ light curves, obtained during 2014 and 2015.} 
\label{table:orbit_act_result}      
\centering                                      
\footnotesize
\begin{tabular}{l c c c  }          
\hline\hline                        
\noalign{\smallskip}
Parameter & Pr\,0211b & Pr\,0211c & Unit    \\    
\noalign{\smallskip}
\hline                                   
\noalign{\smallskip}                                
 P                       & $2.14609 \pm 2 \cdot 10^{-5}$ & $5300^{+4450}_{-1800}$ & days   \\   
 K                        & $309.4 \pm 2.5 $     & $135  \pm 4 $       & \ms   \\
 $\phi$                    & $ 207.0 \pm 0.5 $    & $ 101 \pm 5 $      & deg   \\
 $\sqrt{e} \sin{\omega}$ & $-0.06^{+0.07}_{-0.05}  $    & $ 0.81^{+0.06}_{-0.07} $   &   \\
 $\sqrt{e} \cos{\omega}$  & $ 0.10^{+0.04}_{-0.07}$ &  $ -0.19^{+0.09}_{-0.08} $ &   \\
 e                      \tablefootmark{a}  & $0.017 \pm 0.010 $   & $ 0.70 \pm 0.10$    &  \\
 $\omega $              \tablefootmark{a}  & $329 \pm 35 $        &  $ 103 \pm 6 $      & deg \\
 \mplanet $\sin i$      \tablefootmark{a}  & $1.88 \pm 0.02$  & $7.95 \pm 0.25 $&  \mjup  \\              
a                        \tablefootmark{a} & $0.03184 \pm 0.00015$  & $ 5.8^{+2.9}_{-1.4}  $&  AU \\  
T$_{\rm peri}$             \tablefootmark{a} & $2456678.6 \pm 0.2 $  & $2456709 \pm\ 16$   & days  \\            
\noalign{\smallskip}
\hline
\hline
\noalign{\smallskip}
Parameter                    & HARPS-N & TRES & unit  \\
\hline
\noalign{\smallskip}
$\gamma$             & $ 35035 \pm 8 $    & $128 \pm 7 $ & \ms  \\
jitter RV             & $ 15 \pm 2  $          & $ 13  \pm 7 $ & \ms  \\
jitter RV$_{2013}$  &  -  & $ 19  \pm 9 $ & \ms  \\
 \prot\   & \multicolumn{2}{c}{$7.93 \pm 0.01$}     &  days \\ 
 RV-Phot. $\phi$ & \multicolumn{2}{c}{$263 \pm 4$}   & deg \\   
 T$_{\rm 0}$  \tablefootmark{b} & \multicolumn{2}{c}{2456679.97345}     & days \\    
\hline
\noalign{\smallskip}
\multicolumn{4}{c}{2013 obs. season } \\
\hline
\noalign{\smallskip}
K$_{\rm{P_{rot}}}$    & $50.7 \pm 6.6$  & $44.4 \pm 10.6 $ & \ms \\
K$_{\rm{P_{rot}} /2}$  & $19.2 \pm 5.4$ & -               & \ms \\ 
A$_{\rm{P_{rot}} } $  &  \multicolumn{2}{c}{$0.0251 \pm 0.0005$} & $B$ mag \\
A$_{\rm{P_{rot}} /2}$  &  \multicolumn{2}{c}{$0.0054 \pm 0.0007$}& $B$ mag \\
  $\phi _{\rm{P_{rot}}   }$ &  \multicolumn{2}{c}{ $ 119 \pm 15 $} & deg  \\ 
  $\phi _{\rm{P_{rot}} /2}$ &  \multicolumn{2}{c}{ $ 73 \pm 30 $} & deg  \\ 
offset $B$     &  \multicolumn{2}{c}{$ -0.0098 \pm 0.0004 $  }     & $B$ mag  \\
jitter $B$     &  \multicolumn{2}{c}{$ 0.0033 \pm 0.0003 $  }      & $B$ mag  \\
\noalign{\smallskip}
\hline
\multicolumn{4}{c}{2014 obs. season } \\
\hline
\noalign{\smallskip}
K$_{\rm{P_{rot}}}$    & $27.8 \pm 5.1$  & $15.6 \pm 10.3 $ & \ms \\
K$_{\rm{P_{rot}} /2}$  & $33.8 \pm 5.0$ & -               & \ms \\ 
A$_{\rm{P_{rot}} } $  &  \multicolumn{2}{c}{$0.0246 \pm 0.0008$} & $B$ mag \\
A$_{\rm{P_{rot}} /2}$  &  \multicolumn{2}{c}{$0.0038 \pm 0.0008$}& $B$ mag \\ 
  $\phi _{\rm{P_{rot}}   }$ &  \multicolumn{2}{c}{ $ 215 \pm 7 $} & deg  \\     
  $\phi _{\rm{P_{rot}} /2}$ &  \multicolumn{2}{c}{ $ 322 \pm 10 $} & deg  \\     
offset $B$   &  \multicolumn{2}{c}{$ 0.0044 \pm 0.0006 $}     & $B$ mag  \\
jitter $B_{2014,2015}$   &  \multicolumn{2}{c}{$ 0.0050 \pm 0.0003 $}     & $B$ mag  \\
offset $B_{\rm short}$& \multicolumn{2}{c}{$ 0.0034 \pm 0.0010 $ }      & $B$ mag  \\
jitter $B_{\rm short}$& \multicolumn{2}{c}{$ 0.0082 \pm 0.0008 $ }      & $B$ mag  \\
\noalign{\smallskip}
\hline
\multicolumn{4}{c}{2015 obs. season } \\
\hline
\noalign{\smallskip}
K$_{\rm{P_{rot}}}$    & $32.4 \pm 5.2$  & $28.6 \pm 9.2 $ & \ms \\
K$_{\rm{P_{rot}} /2}$  & $10.4 \pm 5.1$ & -               & \ms \\ 
A$_{\rm{P_{rot}} } $  &  \multicolumn{2}{c}{$0.0124 \pm 0.0008$} & $B$ mag \\
A$_{\rm{P_{rot}} /2}$  &  \multicolumn{2}{c}{$0.0050 \pm 0.0008$}& $B$ mag \\ 
A$_{\rm{P_{rot}} } $  &  \multicolumn{2}{c}{$0.0092 \pm 0.0006$} & $r$ mag \\
A$_{\rm{P_{rot}} /2}$  &  \multicolumn{2}{c}{$0.0035 \pm 0.0005$}& $r$ mag \\ 
  $\phi _{\rm{P_{rot}}   }$ &  \multicolumn{2}{c}{ $ 312 \pm 22 $} & deg  \\        
  $\phi _{\rm{P_{rot}} /2}$ &  \multicolumn{2}{c}{ $  79 \pm 43 $} & deg  \\        
offset $B$   &  \multicolumn{2}{c}{$ 0.0045 \pm 0.0006 $}     & $B$ mag  \\
jitter $B_{2014,2015}$   &  \multicolumn{2}{c}{$ 0.0050 \pm 0.0003 $}     & $B$ mag  \\
offset $r$       &  \multicolumn{2}{c}{$0.0033 \pm 0.0004 $}      & $r$ mag  \\
jitter $r$       &  \multicolumn{2}{c}{$0.0032 \pm 0.0003 $}      & $r$ mag  \\
\noalign{\smallskip}
\hline
\noalign{\smallskip}
\end{tabular}
        \tablefoot{All the parameters have been fitted except:
         \tablefoottext{a}{derived}
         \tablefoottext{b}{fitted}.
         } 
\end{table}

\begin{figure*}[htbp]
\resizebox{\hsize/2}{!}{\includegraphics{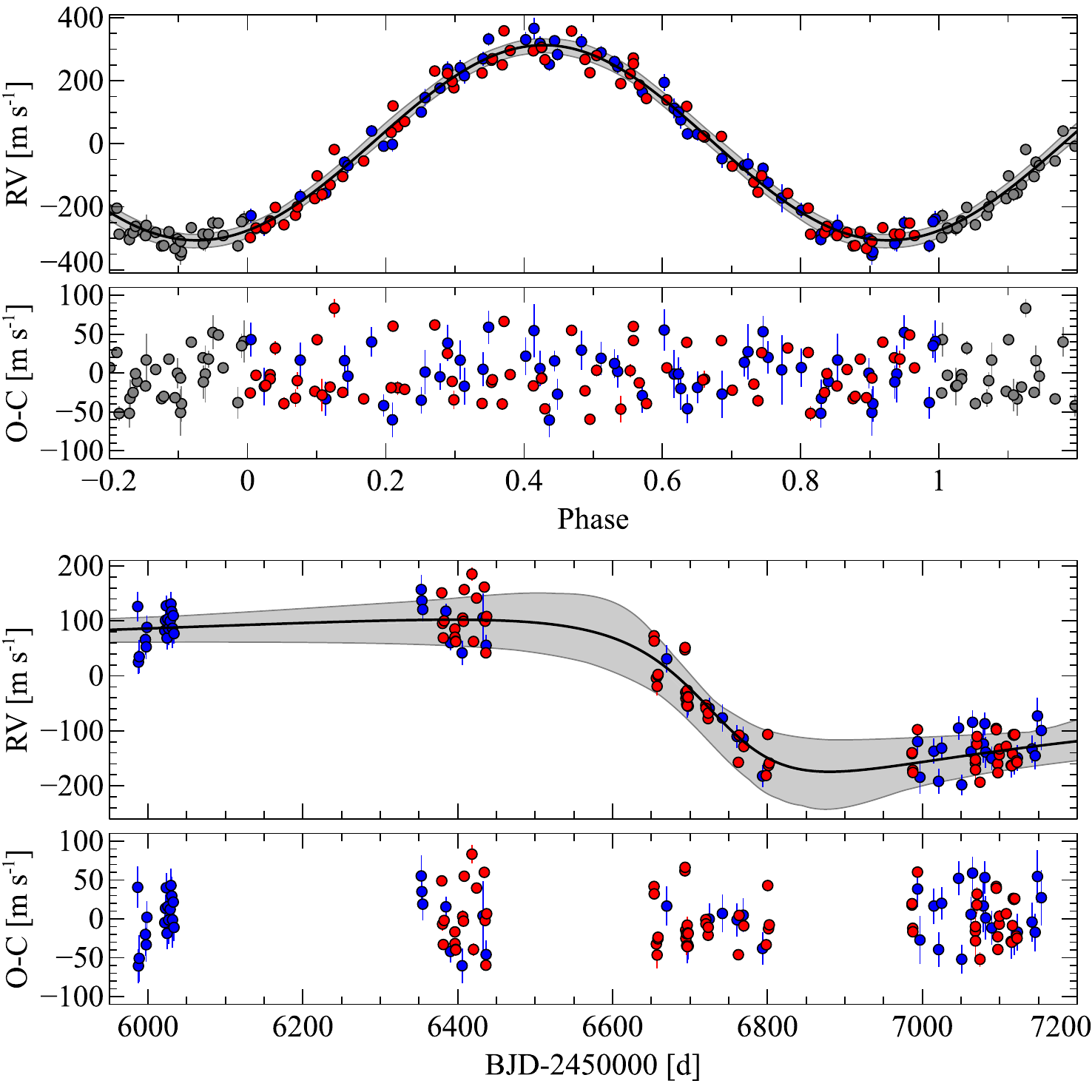}}
\resizebox{\hsize/2}{!}{\includegraphics{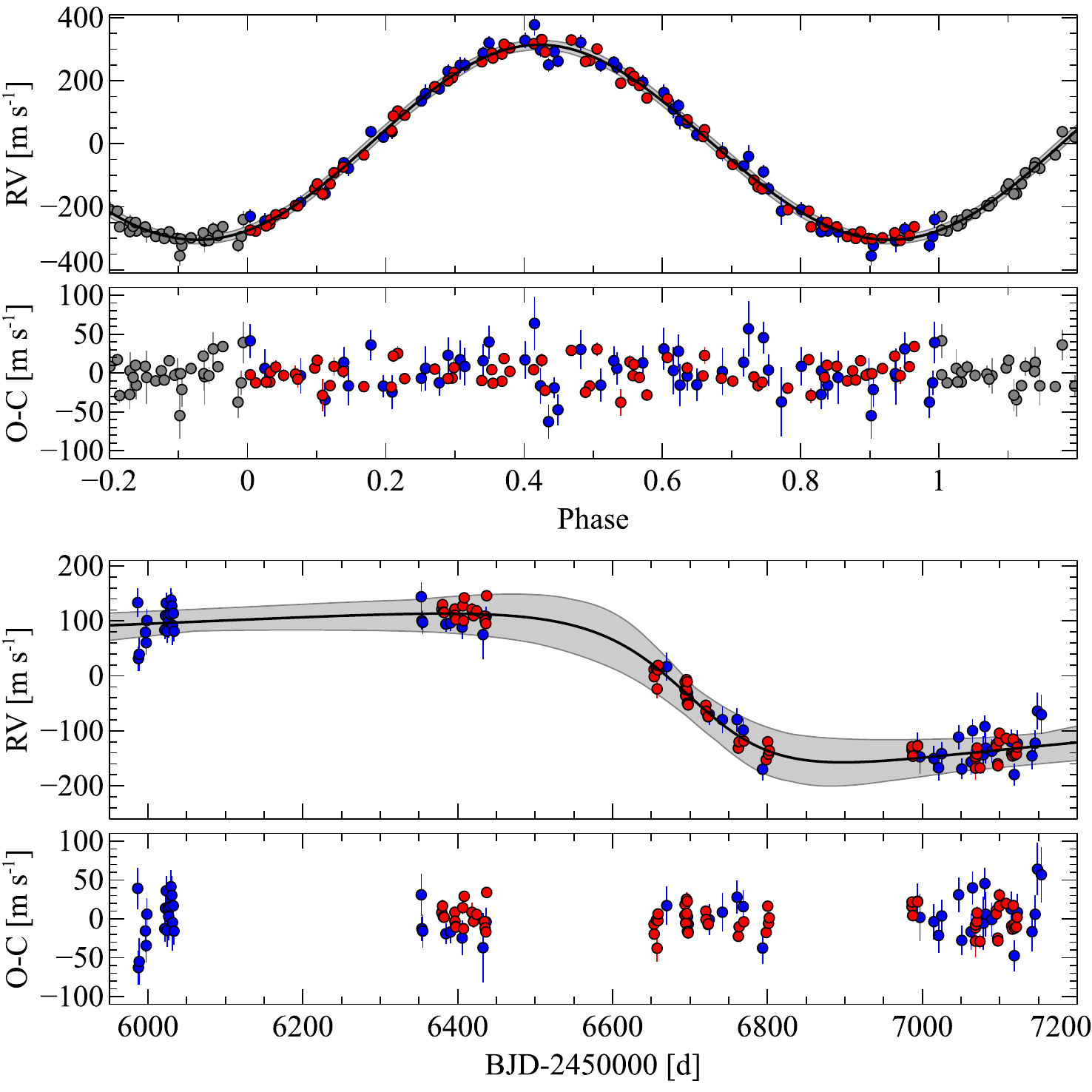}}
\caption{ Orbital solution and RV residuals before (left panels) and after (right panels) the inclusion of activity in the global fit. The two upper panels show the RV fit and its residuals for the inner planet, phased on its period, after removing the solution of the second planet. The two lower panels show the RV fit as a function of time for the outer planet, after removing the solution of the inner one. Red and blue dots represent HARPS-N and TRES data, respectively. The gray shaded areas represent the $3 \sigma$ confidence regions}
\label{fig:rv_orbit_act}
\end{figure*}

\begin{figure}[htbp]
\includegraphics[width=\linewidth]{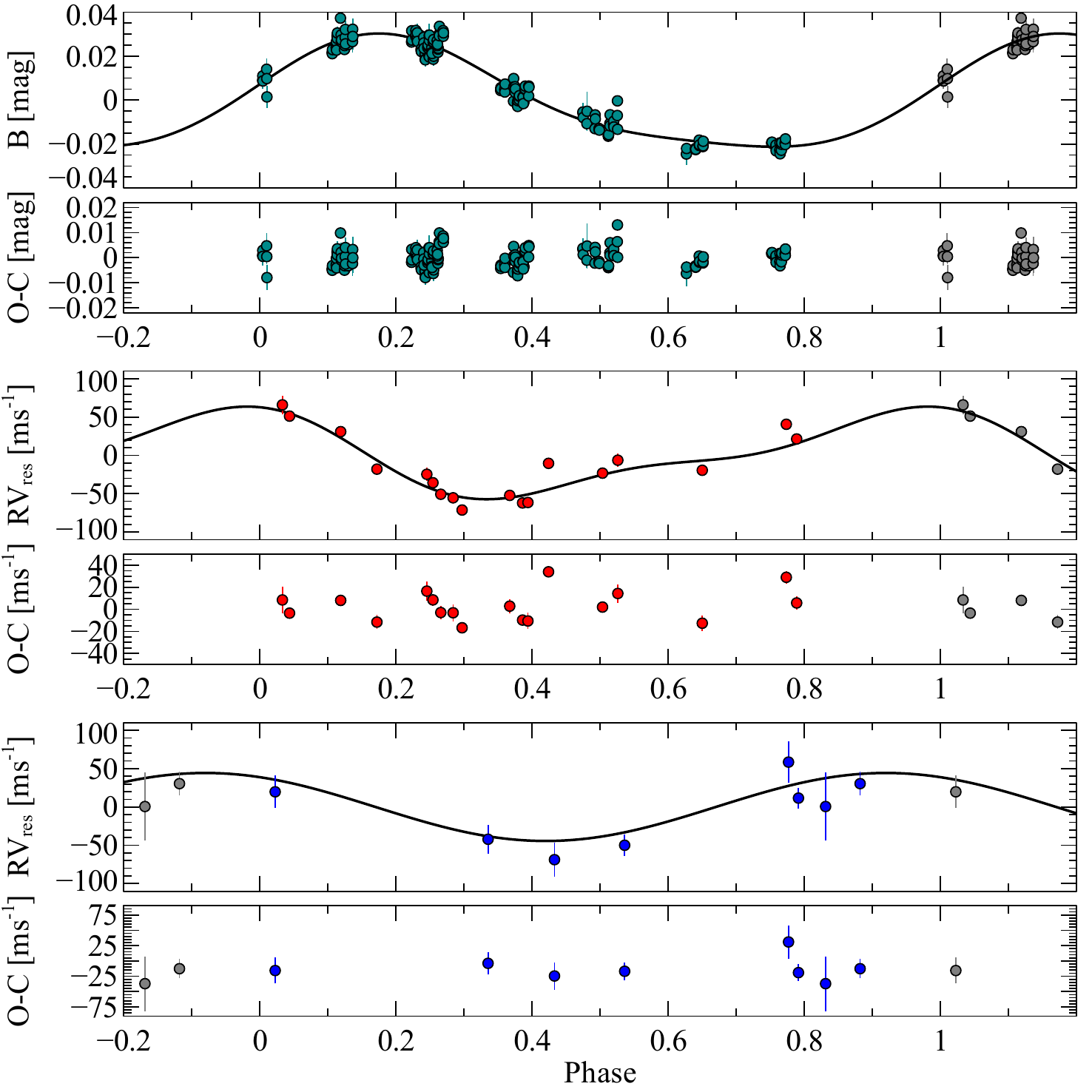}
\caption{Activity model for the data gathered in 2013. The upper panel shows the photometric light curve with residuals. Middle and lower panels show the RV data from HARPS-N and TRES, respectively (color coded as in Figure~\ref{fig:GLS_RV_periods}), after removing the two-planet solution. Data have been phased on the rotational period obtained from the global fit.}
\label{fig:phot_rv_act_P0}
\end{figure}

Our treatment of the activity reduces the errors on  the orbital parameters of the two planets up to $\simeq 30 \%$, as in the case of the semi-amplitudes of the RV signals. The improvement in the RV fit, with a strong reduction of the HARPS-N RV jitter (from 33 \ms\ before to 15 \ms\ after activity correction) and the TRES one (from 26 to 13 \ms ), is visually summarized in Figure~\ref{fig:rv_orbit_act}, where the best-fit model and the RV residuals for the two planets, before and after taking activity into account, are compared side by side. The relationship between the photometric signal and the RVs (after removing the signal of the two planets) is shown in Figure~\ref{fig:phot_rv_act_P0},~\ref{fig:phot_rv_act_P1}, and~\ref{fig:phot_rv_act_P2} for each observational season. The slight loss of coherence of the activity-induced signal in the asymmetry/activity indexes is supported by the change of shape and phase observed in the photometric light curves, as well as the different shapes of the RV sinusoids across the seasons.

\onlfig{
\begin{figure}[htbp]
\includegraphics[width=\linewidth]{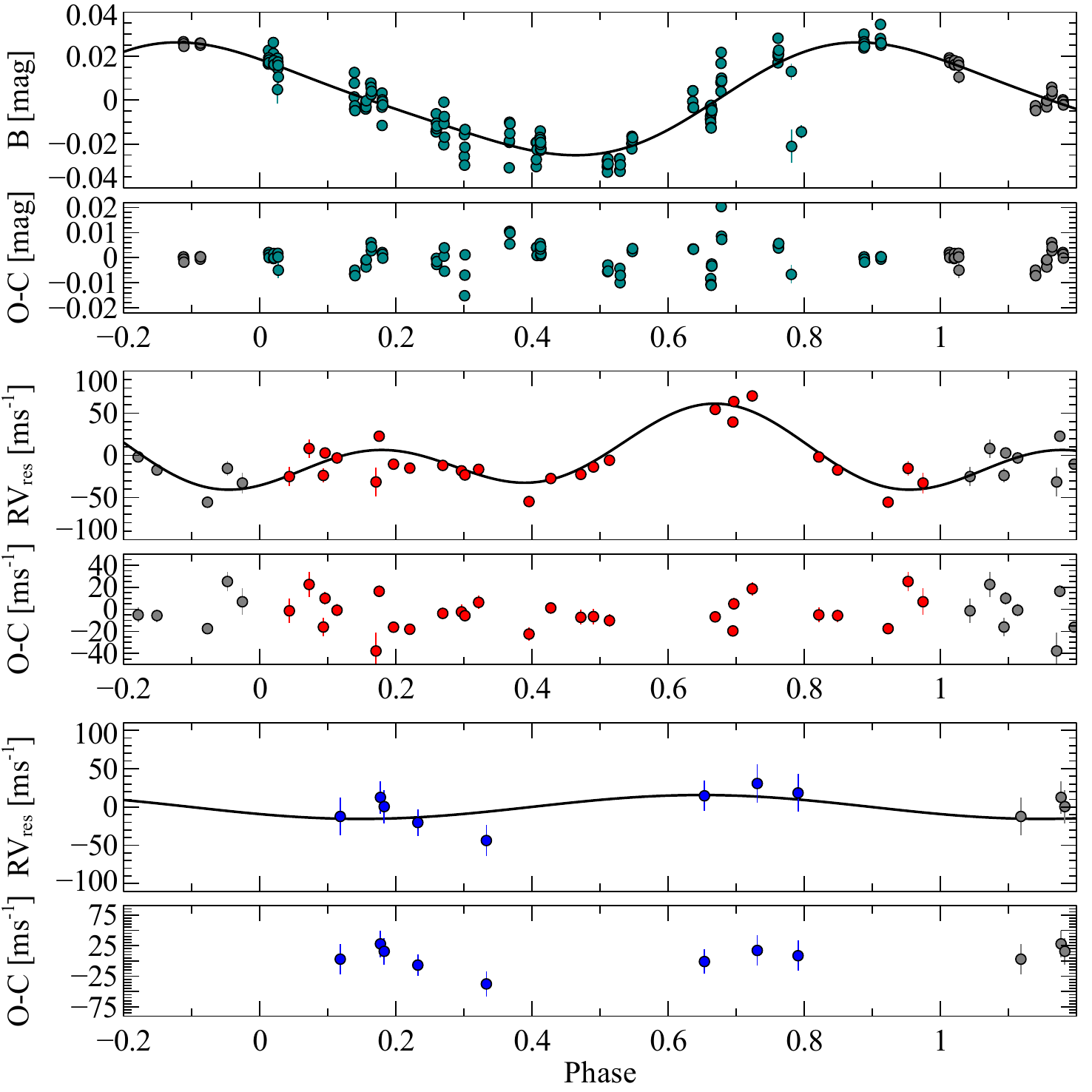}
\caption{As in Figure~\ref{fig:phot_rv_act_P0} but for the 2014 observational season.}
\label{fig:phot_rv_act_P1}
\end{figure}
}

\onlfig{
\begin{figure}[htbp]
\includegraphics[width=\linewidth]{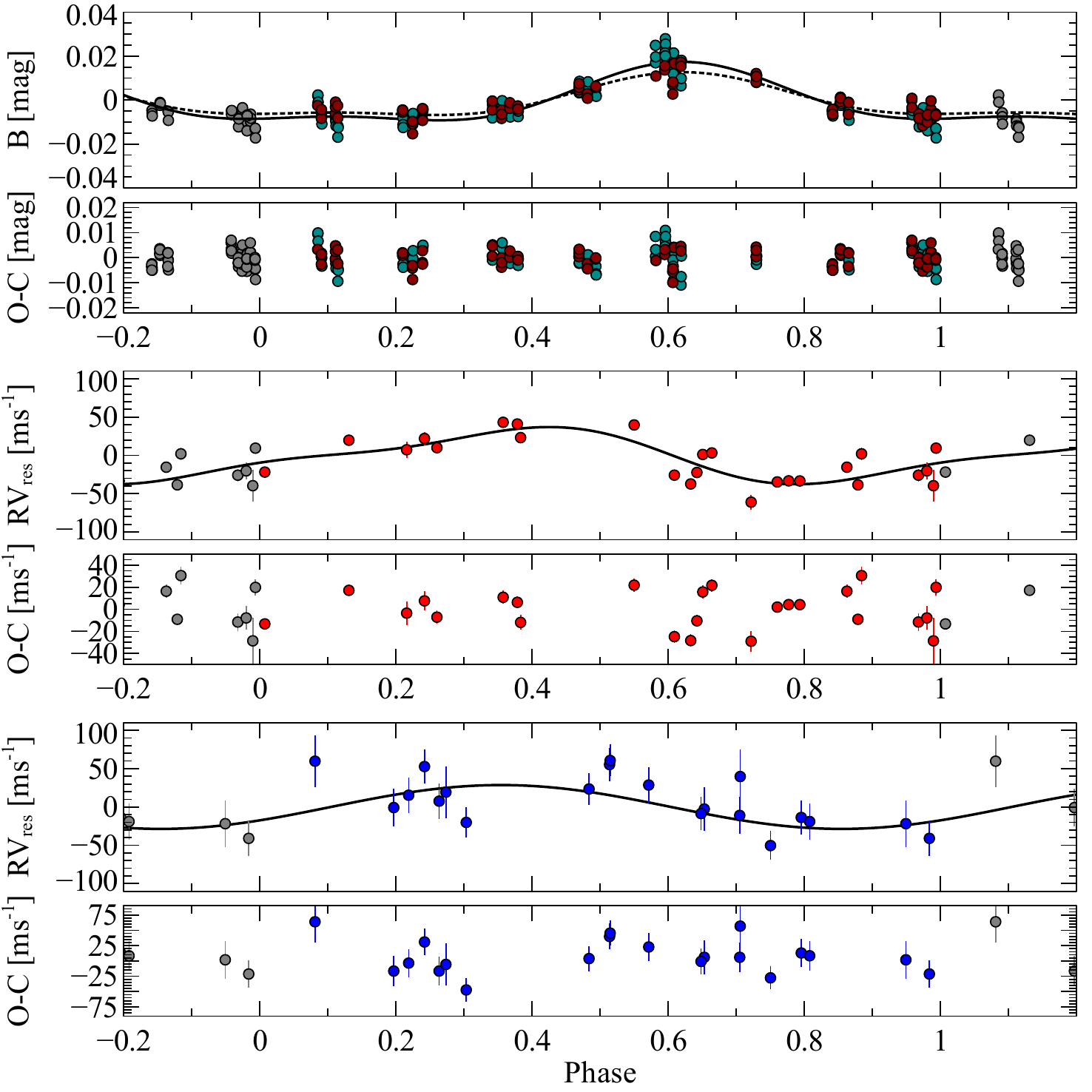}
\caption{As in Figure~\ref{fig:phot_rv_act_P0} but for the 2015 observational season. Data in the $r$ filter in the photometric light curve are represented by dark red points, while the activity model in this band is marked with a dotted line.}
\label{fig:phot_rv_act_P2}
\end{figure}
}

The median values of the orbital parameters do not change significantly after including the activity model in the overall fit. This implies that the two observed planetary signals are not an artifact that was introduced by our methodology to correct for the  activity-induced RV variations. However, we note that this may not hold in general, and that Pr\,0211 represents a special case because of the large separation in frequency space between the planetary and activity signals. 

With an eccentricity of $0.02 \pm 0.01$, following the criterion of \cite{Lucy1971}, the inner planet can be safely considered as being on a circular orbit. Our determination improves the previous determination of Q12 by a factor of two ($e = 0.05 \pm 0.02$).

\begin{figure}[htbp]
\includegraphics[width=\linewidth]{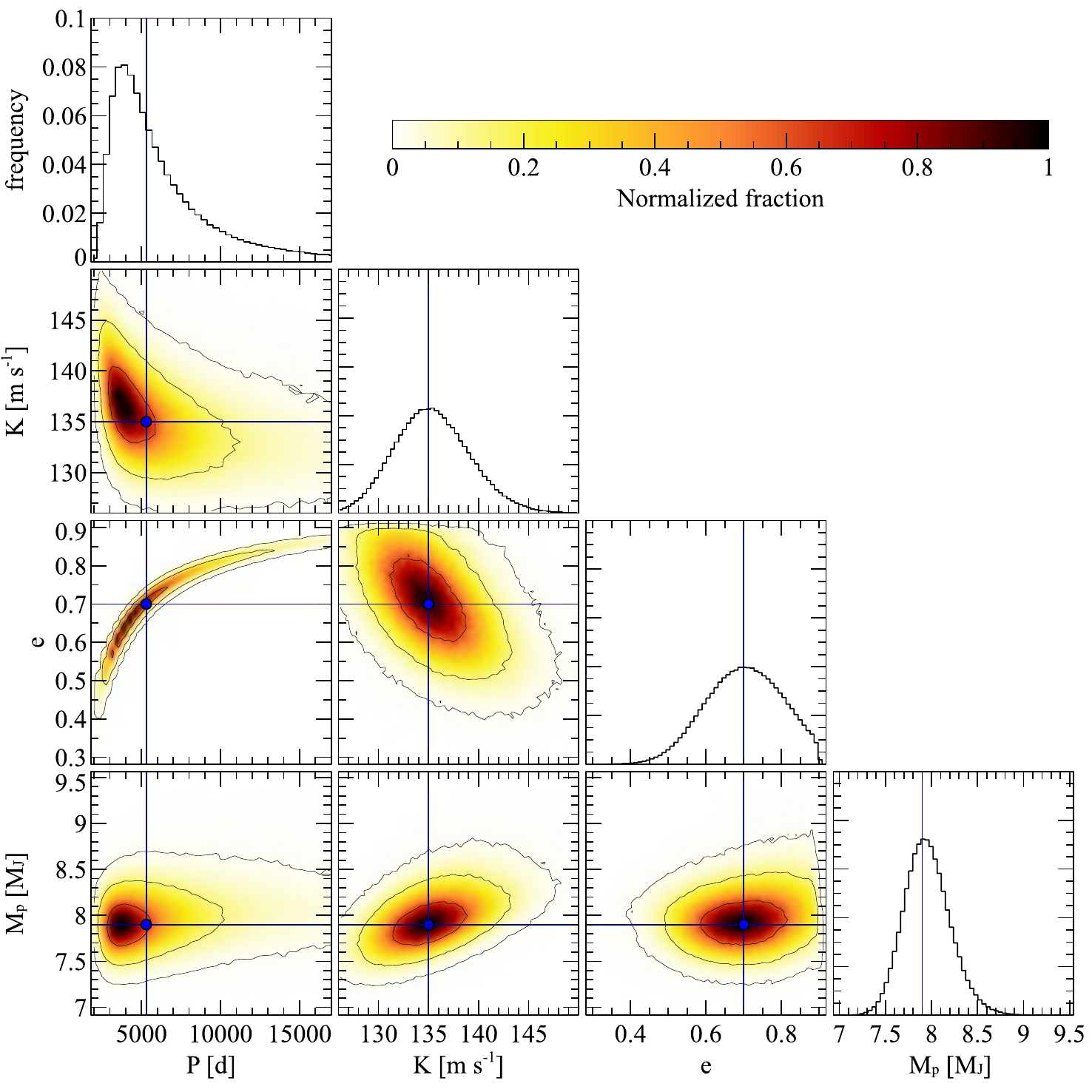}
\caption{Posterior distribution of period $P$, RV semi-amplitude $K$, orbital phase, eccentricity $e$, and planetary mass for Pr\,0211c. Histograms share the same vertical scale. Blue crosses identify the median values reported in Table~\ref{table:orbit_act_result}.}
\label{fig:rv_planetc_posterior}
\end{figure}

The limited time-span of our observation allows us to only put a limit on the period of the outer between $9.6$ and $27$ years, and between $0.6$ and $0.8$ on its eccentricity, with a strong degeneracy between the two parameters. The projected mass range is well constrained at $ \simeq 7.95 \pm 0.25$ \mjup\ thanks to the observation of both the turning points in the RV curve. The relationship between these parameters is visualized in the density plot of the posterior distribution for Pr\,0211c, in Figure~\ref{fig:rv_planetc_posterior}.

RV residuals do not show any significant peak in their periodogram (Figure~\ref{fig:rv_residuals_GLS_k4}). No additional signals that could be interpreted as planets are detected in our data. We can safely rule out the presence of a planet with an RV semi-amplitude greater than the observed RV jitter.

\onlfig{
\begin{figure}[htbp]
\includegraphics[width=\linewidth]{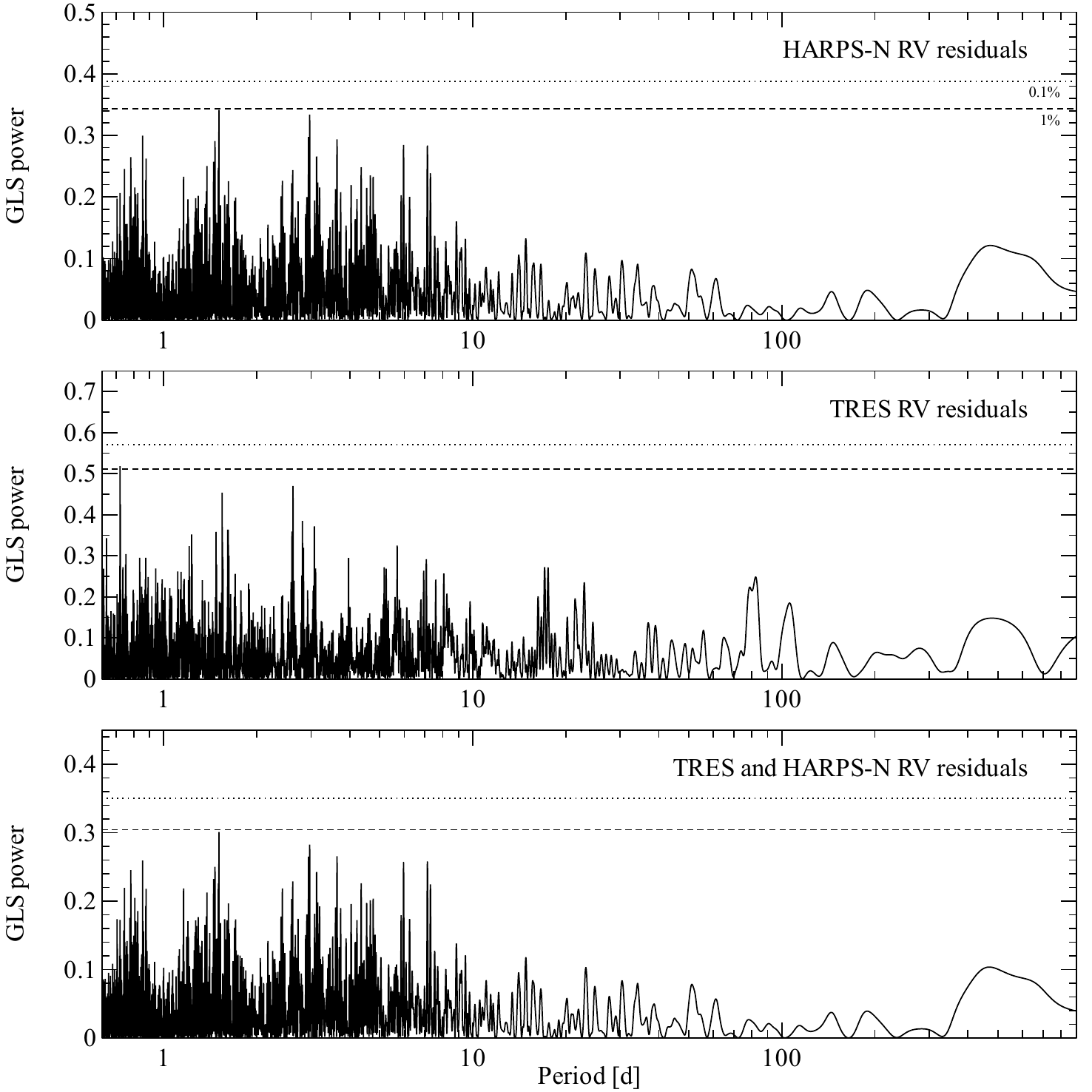}
\caption{ GLS periodograms of the RV residuals after subtracting the 2-planets solution and the activity model. In the case of TRES we have included only RVs in the temporal range covered by the activity model, \ie , data gathered since 2013.}
\label{fig:rv_residuals_GLS_k4}
\end{figure}
}

\section{Discussion}\label{sec:discussion}
The activity of Pr\,0211 is noticeable if compared to the average RV jitter of 13 \ms\ found by Q12 for this cluster, and 16 \ms\ found by \cite{Paulson2004} for the almost coeval Hyades cluster. For this star, photometric sinusoidal variations correspond to RV amplitudes up to 50 \ms , and the RV jitter term in the combined fit of the planet motion and activity signal for the HARPS-N RVs is still very high ($\simeq 15 $ \ms ), compared to the average internal error of the RVs ($\simeq 6 $ \ms ). One possibility could be that the star is subjected to star-planet interactions caused either by the inner planet, which could explain the short-term variability, but the low \snr\ of the spectra does not allow us to draw any conclusions on the possible connection between the two planets and their host star. The photometric amplitude of $\simeq 0.025$ mag that we measure in the B filter during the first two seasons is higher than the 0.017 mag measured in white light by K14. This is expected since at shorter wavelengths the contrast between the cool stellar spots and the hotter surrounding photosphere is larger.

The dynamical stability of the system has been verified by running a 100 Myr simulation with {\tt SyMBA} \citep{Duncan1998}. We used the values listed in Table \ref{table:orbit_act_result} as initial parameters for the planets, performing three separate analyses with the median values and their $\pm \sigma$ variations for the orbital parameters of Pr\,0211c, and the median parameters for Pr\,0211b.

We also estimated the characteristic tidal timescales of the host star and the inner planet using the model of \cite{Leconte2010} and assuming the stellar parameters in Table~\ref{table:stellar_parameters}, the planetary parameters in Table~\ref{table:orbit_act_result},and a radius of 1.0 \rjup\ for the planet. 
We assumed modified tidal quality factors $Q^{\prime}_{\rm s} =10^{7}$ for the star \citep{Jackson2009}{\bf ,} and $Q^{\prime}_{\rm p} = 10^{5}$ for the planet, similar to the value  measured for Jupiter \citep{Lainey2009}. The timescale for the circularization of the orbit is ruled by the dissipation inside the planet and it is much shorter ($\sim 15$~Myr) than the age of the system, thus supporting our result about a circular orbit in Section~\ref{sec:activity_orbit}. The timescales of tidal orbital decay and star spin alignment are ruled by the tidal dissipation inside the star and are proportional to $Q^{\prime}_{\rm star}$. With the adopted value and the present system parameters, they are of $\approx 6$ and $\approx 11$~Gyr respectively, thus significantly longer than the age of the star. Therefore, a possible primordial misalignment between the stellar spin and the orbital angular momentum could still be present.

While the analysis of Kepler candidates seems to suggest that HJs do not generally have small-size companions (Steffen et al. 2009), 
\nocite{Steffen2012}
several examples of HJs with periods shorter than 10 days and outer massive companions with P $> 100$ do exist. For example, the planetary systems around $\upsilon$ And (Butler et al. 1997), 
\nocite{Butler1997}
HD\,187123 (Wright et al. 2007),
\nocite{Wright2007}
HAT-P-13 (Bakos et al. 2009),
\nocite{Bakos2009}
HD\,217107 and HIP\,14810 \citep{Wright2009}, Kepler-424 \citep{Endl2014}, WASP-41 and WASP-47 \citep{Neveu-VanMalle2015}. Among these systems, only HAT-P-13 and HD\,217107 show a similar architecture to Pr\,0211, \ie , the outer planet has high eccentricity $(e > 0.5)$. A recent investigation by \cite{Knutson2014}, which was aimed at finding outer companions to HJs, estimated a frequency of $50 \pm 10$ \% 
outer planets with mass 1-13 \mjup\ within 1$-$10 AU from their hosting star. In fact, the observed statistics on the number of HJs with outer planets is strongly affected by several biases, both because the host star is usually excluded from the RV surveys after the discovery of an HJ, and because of the difficulty of disentangling long-term magnetic cycles from planetary signals (\eg , \citealt{Damasso2015}).

The present orbital architecture of Pr\,0211 may be typical of a planetary system with at least three planets, which has experienced a period of chaotic dynamics evolving into planet-planet scattering. Eventually, two planets were left on stable, inner orbits, while the other ones are ejected from the system on a hyperbolic trajectory.
Subsequent tidal interaction with the host star circularize the orbit of the inner planet to the periastron distance, transforming it into a HJ, while the outer one survives on an eccentric and misaligned orbit \citep{Weidenschilling1996,Chatterjee2008,Nagasawa2008}. 

In this scenario, two viable processes triggered the planet-planet scattering phase: the planets have been on dynamically unstable orbits ever since their formation, or the planetary system experienced  a stellar encounter in the initial evolution of the cluster.
In the first case, the fraction of planetary systems in clusters that underwent a planet-planet scattering process would be comparable to that of field stars or only slightly higher. In the second case, since close stellar flybys are frequent in the initial stages of cluster evolution, the typical orbital architecture of planets around stars in clusters is expected to be shaped by planet-planet scattering events more frequently than for field stars \citep{Zakamska2004,Malmberg2009}. As a consequence, we should have a large number of systems with a HJ and a second giant planet on an outer eccentric orbit with respect to the second case. Recent simulations have tried to tackle the problem by determining the expected frequency of such systems \citep{Hao2013,Li2015} or advocating a cluster origin for several field stars with planetary systems similar to that of Pr\,0211 \citep{Shara2014arXiv}.

On the other hand, the majority of cluster stars  disperse into the field within the first 10 Myr (\eg , \citealt{Fall2009,Dukes2012}), so stellar encounters are more probable during the early planetary formation phase when the star is still embedded in its nest disc. 
In this case, the dynamical excitation of a stellar encounter is fully damped by the circumstellar disc and any trace of the flyby is erased \citep{Marzari2013,Picogna2014}. The probability of stellar encounters at later time, however, could be increased if the cluster is the outcome of two smaller clusters merging, as it has been suggested for M\,44 \citep{Holland2000}.

More planetary systems in open clusters are needed to clarify the contribution of stellar flybys to the final orbital configurations of multi-planet systems that were originally formed in clusters.

\section{Conclusion}\label{sec:conclusion}

We have presented the discovery of the first planetary system around an open cluster star.
New HARPS-N and TRES radial velocity measurements of the planet-host star Pr\,0211 in M\,44 have led us to the detection of a long-period,  massive Jupiter and to confirm the nearly-circular orbit of the already known inner planet. A deep analysis of the stellar activity has been performed to exclude a stellar cause for the additional RV signal. We then performed a joint modeling of the activity and the planetary signals to improve the precision of the orbital parameters of the two planets, with the inclusion of a photometric light curve that had been specifically gathered for this purpose. 

The discovery of the first multi-planet system in an open cluster is the first result of a long-term search for planets in OCs, which is conducted within the GAPS collaboration, and which will be described in a forthcoming paper.


\begin{acknowledgements}
GAPS acknowledges support from INAF through the ``Progetti Premiali'' funding scheme of the Italian Ministry of Education, University, and Research.
LM acknowledges the financial support provided by the European Union Seventh Framework Programme (FP7/2007-2013) under Grant agreement number 313014 (ETAEARTH).
VN acknowledges partial support from INAF-OAPd through
the grant ``Analysis of HARPS-N data in the framework of GAPS project''
(\#19/2013) and ``Studio preparatorio per le osservazioni della 
missione ESA/CHEOPS'' (\#42/2013).
JIGH acknowledges financial support from the Spanish Ministry of Economy and Competitiveness (MINECO) under the 2013 Ramón y Cajal program MINECO RYC-2013-14875, and the Spanish ministry project MINECO AYA2014-56359-P.
We made use of the SIMBAD database and VizieR catalogue access tool, operated at the CDS, Strasbourg, France.
We wish to thank the anonymous referee for the valuable comments and suggestions.
\end{acknowledgements}


\bibliographystyle{aa} 
\bibliography{Bibliography} 

\end{document}